# A 500 pc volume-limited sample of hot subluminous stars

## I. Space density, scale height, and population properties

H. Dawson[1], S. Geier[1], U. Heber[2], I. Pelisoli[3,1], M. Dorsch[1,2], V. Schaffenroth[4,1], N. Reindl[5,1], R. Culpan[1], M. Pritzkuleit[1], J. Vos[8,1], A. A. Soemitro[1,15], M. M. Roth[1,15], D. Schneider[2], M. Uzundag[19,6,7], M. Vučković[6], L. Antunes Amaral[6,7], A. G. Istrate[9], S. Justham[10,24], R. H. Østensen[11], Telting, J. H.[12,21], A. A. Djupvik[12,21], R, Raddi[13], E. M. Green[14], C. S. Jeffery[16], S. O. Kepler[17], J. Munday[3,20], T. Steinmetz[20,22,23], and T. Kupfer[25]

*(Affiliations can be found after the references)*

**ABSTRACT**

We present the first volume-limited sample of spectroscopically confirmed hot subluminous stars out to 500 pc, defined using the accurate parallax measurements from the *Gaia* space mission data release 3 (DR3). The sample comprises a total of 397 members, with 305 (∼ 77%) identified as hot subdwarf stars, including 83 newly discovered systems. Of these, we observe that 178 (∼ 58%) are hydrogen-rich sdBs, 65 are sdOBs (∼ 21%), 32 are sdOs (∼ 11%), and 30 are He-sdO/Bs (∼ 10%). Among them, 48 (∼ 16%) exhibit an infrared excess in their spectral energy distribution fits, suggesting a composite binary system. The hot subdwarf population is estimated to be 90% complete, assuming that most missing systems are these composite binaries located within the main sequence (MS) in the *Gaia* colour-magnitude diagram (CMD). The remaining sources in the sample include cataclysmic variables (CVs), blue horizontal branch stars (BHBs), hot white dwarfs (WDs), and MS stars. We derived the mid-plane density $\rho_0$ and scale height $h_z$ for the non-composite hot subdwarf star population using a hyperbolic sechant profile (sech$^2$). The best-fit values are $\rho_0 = 5.17 \pm 0.33 \times 10^{-7}$ stars/pc$^3$ and $h_z = 281 \pm 62$ pc. When accounting for the composite-colour hot subdwarfs and their estimated completeness, the mid-plane density increases to $\rho_0 = 6.15^{+1.16}_{-0.53} \times 10^{-7}$ stars/pc$^3$. This corrected space density is an order of magnitude lower than predicted by population synthesis studies, supporting previous observational estimates.

**Key words.** stars: subdwarfs – catalogs – stars: binaries – stars: Hertzsprung-Russell and colour-magnitude diagrams – stars: statistics

## 1. Introduction

Since their serendipitous discovery by Humason & Zwicky (1947) as faint blue stars at high galactic latitudes, hot subluminous stars have been characterised by their position in the Hertzsprung–Russell Diagram (HRD) as having lower intrinsic luminosities compared to early-type main sequence (MS) stars of similar colour. This region of the HRD between the MS and the white dwarfs (WDs) hosts many types of stars, each with its own associated evolutionary path. The hot subdwarf stars of spectral type B and O (sdBs and sdOs) compose a prominent subgroup of this population. They have since been associated with, although not restricted to, the extreme horizontal branch (EHB, Greenstein & Sargent 1974; Newell & Sadler 1978; Heber et al. 1984) and the helium main sequence (Paczyński 1971), or have evolved beyond these stages to even higher temperatures. However, to qualify for the EHB in the canonical sense, the progenitor must be stripped of nearly its entire hydrogen envelope during core He burning. Since the majority of hot subdwarfs have masses close to the core-helium-flash mass of ≈ 0.47 $M_\odot$ (Fontaine et al. 2012; Schaffenroth et al. 2022), stripping is often associated with the tip of the red giant branch (RGB) for low-mass stars with degenerate cores (Dorman & Rood 1993). However, since an intermediate mass (2.3 − 8 $M_\odot$) possible red giant progenitor may ignite helium in its non-degenerate core at an earlier stage, hot subdwarf stars are not restricted to this specific final mass (Han et al. 2002; Hu et al. 2008; Prada Moroni & Straniero 2009; Götberg et al. 2018). Regardless, the result is a hotter and more compact object than canonical horizontal branch (HB) stars with surface gravity and effective temperature ranges of $T_{\text{eff}} = 20\,000 − 40\,000$ K and log $g$ = 4.5 - 6.2 dex, and with radii between 0.15 $R_\odot$ and 0.35 $R_\odot$ (see Heber 2016, for a complete review). Consequently, they will not ascend the asymptotic giant branch (AGB), but instead will evolve directly towards the WD cooling track.

The large amount of mass loss required to form EHB stars at the point of He burning is difficult to explain in the context of single-star evolution, and remains a missing piece of the puzzle in stellar evolution theory. Although single-star scenarios are still discussed (Sweigart 1997; Castellani & Castellani 1993; Miller Bertolami et al. 2008), decades of observational research have revealed a high binary fraction in these stars, which either host compact companions like WDs or low-mass MS stars in close binaries (Maxted et al. 2001; Napiwotzki et al. 2004; Geier et al. 2022; Schaffenroth et al. 2022, 2023) or cool MS stars in wide binaries (Stark & Wade 2003; Barlow et al. 2012, 2013; Vos et al. 2012, 2013), or otherwise have provided strong evidence of binary interaction (Pelisoli et al. 2020). This shifted the focus to binary evolution scenarios invoking three main formation channels regarding common-envelope evolution (CEE, Paczynski 1976), Roche-lobe overflow (RLOF, Han et al. 2002; Chen et al. 2013), and mergers (e.g. He-WD + He-WD mergers (Webbink 1984), CEE mergers (Politano et al. 2008), hybrid mergers (Justham et al. 2011), and He-WD + low-mass MS mergers (Clausen & Wade 2011)), the last of these may explain the observed population of single hot subdwarf stars (e.g. Napiwotzki et al. 2004; Geier et al. 2022). This motivated Han et al. (2002, 2003), and later Clausen et al. (2012), to perform a detailed binary population synthesis (BPS) study of the formation of EHB stars. The primary objective of these studies was to discern the relative importance of the three main evolutionary channels leading to their formation, and how different combinations of input parameters can alter the resulting population properties. A key statistical parameter derived in Han et al. (2002, 2003) is the space number density, $\rho$, which was found to be $\rho = 1 \times 10^{-5}$ stars/pc$^3$ and corresponds to a total of 2.4–9.5 million hot subdwarf stars in the Galaxy.

Kilkenny et al. (1988) provided the first catalogue of spectroscopically identified hot subdwarfs which totalled 1225 sdO/Bs. Østensen (2004) compiled the first online database of more than 2300 entries largely based on surveys targeting extragalactic sources (Hagen et al. 1995; Wisotzki et al. 1996; Mickaelian et al. 2007; Mickaelian 2008). Since then, large-scale sky surveys like the Sloan Digital Sky Survey (SDSS) (Geier et al. 2015; Kepler et al. 2015, 2016, 2019), Edinburgh-Cape (EC) survey (Stobie et al. 1997), and the Galaxy Evolution Explorer (GALEX) (Vennes et al. 2011) have contributed extensively to this number using both spectroscopy and photometry. In an effort to provide an up-to-date catalogue of hot subdwarf stars, Geier et al. (2017) compiled 5613 unique sources whose classifications are based on spectroscopy and photometry from these large-scale surveys, as well as radial velocities, atmospheric parameters, and light-curve variations where available. This list has since been updated (Geier 2020; Culpan et al. 2022) in the light of the copious all-sky survey data now available, adding over 500 more objects to the catalogue of known hot subdwarfs in our Galaxy.

The main motivation behind compiling catalogues of hot subdwarf stars is to constrain evolutionary models. In this context, early determinations of their spatial distribution, based on flux-limited samples, yielded conflicting scale heights from 175 to ≈1,000 pc. This discrepancy extended to the derived space densities, varying between ≈$3.8 \times 10^{-7}$ to $4 \times 10^{-6}$ stars/pc$^3$ (see Table 1 for an overview of the previous studies). In short, these previous investigations focused on specific and limited sky regions, such as the poles (Heber 1986; Saffer 1991), Galactic disk (Downes 1986; Villeneuve et al. 1995b), and intermediate latitudes (Moehler et al. 1990a; Theissen et al. 1993; Saffer 1991) of the Galaxy. Villeneuve et al. (1995a) used the entire footprint of the Palomar Green (PG) survey (Green et al. 1986). These studies, constrained by flux-limited samples, faced limitations in the maximum distances and volumes that could be studied, with notable incompleteness at bright magnitudes and misclassifications. These challenges were identified as major caveats, as pointed out by Moehler et al. (1990a) and thoroughly discussed by Villeneuve et al. (1995a). The deepest samples primarily included stars at distances typical for thick-disk stars, suggesting potential incompleteness in modern catalogues of known hot subdwarfs (Geier et al. 2017; Geier 2020; Culpan et al. 2022) at bright magnitudes. Consequently, there is a need for an all-sky survey with precise and established completeness.

Complete volume-limited samples offer ideal benchmarks for population studies with their reduced selection effects as they can be directly compared to the output of a population synthesis code. The first attempt at defining a volume-limited sample of hot subdwarf stars, to the best of our knowledge, was Stark & Wade (2003) who, using photometric data from the Two Micron All Sky Survey (2MASS) second incremental data release catalogue, corrected for the selection bias by removing those hot subdwarfs that they observed to be composite, but would otherwise not have been observed if they had been single objects. However, the sample remains to some extent magnitude-limited as no definitive volume can be drawn in this way.

Parallax measurements offer the most direct method for determining distances to nearby stars. Crucially, the *Gaia* space observatory has significantly improved the accuracy of these measurements across all declinations, allowing for the selection and cross-matching of complete all-sky samples with abundant spectroscopic data from the literature (e.g. Gaia Collaboration et al. 2021; Kilic et al. 2020). In its Early Data Release 3 (Gaia Collaboration 2020), the accuracy of astrometry was significantly improved by the extended time baseline since the previous groundbreaking Data Release 2 (DR2) (Gaia Collaboration et al. 2018a). Leveraging this progress, Culpan et al. (2022) (hereafter CG22) curated a catalogue of hot subluminous star candidates. The selection was based on their positions in the colour-absolute magnitude parameter space in the HRD, primarily confined between main sequence stars of spectral types O and B and white dwarfs (Østensen 2006; Geier et al. 2017; Geier 2020). Moreover, the authors generated a main sequence rejection criterion to remove the millions of main sequence stars located redwards of this cut that would contaminate the sample (see Fig. 4 of CG22). Ultimately, the good parallax selected catalogue contains 13,123 candidates from which our volume-limited sample is directly drawn. In this paper we consider stars at distances closer than 500 pc as indicated by their Gaia DR3 parallaxes, which roughly corresponds to a magnitude limit of $m_G = 12 - 14$ mag for unreddened samples, given the spread of absolute magnitudes of sdB stars (see Fig. 11). We note that fewer than 50 stars lying within 500 pc (see Table 1) were included in the early investigations, while the sample studied here is about seven times larger. This offers, for the first time, an unbiased sample which we utilise to derive the most precise estimates of the local space density and spatial number distributions of hot subdwarf stars to date.

## 2. Sample selection

### 2.1. Defining the distance limit, known candidates, and possible chance alignments

Utilising the CG22 catalogue of good parallax candidates, we began by selecting a suitable volume for a statistical study of the population. We were aiming for a balance between a sample size large enough for robust statistical analysis and observational feasibility within a few years. Choosing a volume defined by 500 pc effectively encompasses the local Galactic disk centred on the Sun. We applied a parallax threshold, accounting for the zero-point correction, as follows:

$$\omega_{zp} + 1\sigma_\omega \geq 2 \text{ mas}. \quad (1)$$

Here $\omega_{zp}$ is the zero-point corrected parallax and $\sigma_\omega$ is the standard error on that parallax; moreover, the parallax error takes into account the inflation factor outlined in El-Badry et al. (2021). The above corresponds to a distance limit of $d_\omega \leq 500 + 1\sigma_{dist}$ pc (i.e. no farther than the distance uncertainty) and results in 584 candidate sources. The mean fractional parallax error of all sources in our sample is 5.6% which is well below the 20% threshold limit given in Bailer-Jones (2015), Bailer-Jones et al. (2018), and Bailer-Jones et al. (2021). Accordingly, we appropriate the inversion of parallaxes for all stars in this sample and adopt no prior information to improve the distance measurements.

Following this selection defined by Eq. 1, we performed an external cross-match with the latest catalogue of spectroscopically identified hot subdwarf stars (Geier 2020; Culpan et al. 2022) using TOPCAT (Taylor 2005); all 192 previously known sources within 500 pc were in our sample. The literature search was conducted using the SIMBAD Astronomical Database[1] (Wenger et al. 2000), identifying a total of 260 out of the remaining 392 candidates. These included CVs, WDs, and MS stars, all listed in Table A.3 in the appendix. However, as hot subdwarfs are notoriously misclassified as main sequence stars of spectral types O, B, or A, those without accessible or poor quality spectra in the literature were subject to follow-up spectroscopic observation.

| Survey & study | Field size (deg²) | Spectral type | Brightness limit (mag)[a] | Number of stars | Within 500 pc (Gaia DR3) | Scale height (pc) | Space density ($10^{-7}$ stars pc$^{-3}$) |
|---|---|---|---|---|---|---|---|
| SB[*] (Heber 1986) | 840 | sdB | 14.2(y/V) | 12 | 6 | 190–220 | 40 |
| KPD[†] (Downes 1986) | 1,144 | sdB/sdO | 15.3(B) | 31/20 | 1 | 175 | 20/7 |
| KPD (Villeneuve et al. 1995b) | 1,144 | sdB/sdO (H-rich) | 15.0(V) | 25 | 1 | | $3.8 \pm 1.7$ |
| PG[‡] (Moehler et al. 1990a) | 712 | sdB | 14.2(y) | 11 | 5[b] | 250 | 10 |
| PG (Theissen et al. 1993) | 600 | sdB | 14.2(y) | 11 | 3[b] | $180^{+190}_{-60}$ | $19^{+32.4}_{-13.4}$ |
| PG (Saffer 1991; Saffer & Liebert 1995) | $3 \times 1,200$ | sdB | 15.4(y/V) | 68 | 19[b] | $285^{+120}_{-35}$ | 7.5 |
| PG (Villeneuve et al. 1995a) | 10,714 | sdB/sdO (H-rich) | 15(y) | 209 | ≈ 40 | $600 \pm 150$ | $3 \pm 1$ |
| This work | All-sky | All considered | Volume-limited | 305 | 305 | $281 \pm 62$ | $6.15^{+1.16}_{-0.53}$ |

[*] Slettebak & Brundage (1971)
[†] Downes (1986)
[‡] Green et al. (1986)
[a] Photometric magnitudes: Strømgren *y*, Johnson *B*, *V*
[b] Stars also in the sample of Villeneuve et al. (1995a)

Table 1: Previous population studies of hot subdwarf stars

*Gaia* has the angular resolution to distinguish sources on the sky whose fainter component would otherwise be completely blended in other spectroscopic and photometric observations. TYC8357-3863-1 is one such case identified where the proximity of HD 165493, an apparently nearby bright star, has hindered the acquisition of a spectrum. Coronagraphy may assist in isolating this system. No other chance alignment was identified in this sample.

### 2.2. Further cleaning

The release of *Gaia* DR3 has provided new astrometric quality control parameters that enable sources with unreliable astrometric solutions to be identified. The *astrometric_excess_noise* (AEN) (see Lindegren et al. 2012, for details on the astrometric modelling) and the re-normalised unit weight error (RUWE)[2] are two such commonly used parameters and are recommended by the *Gaia* Data Processing and Analysis Consortium (DPAC) to be $< 1.0$ and $< 1.4$, respectively.

We find that these thresholds are not appropriate in our selection, as many known hot subdwarf stars drastically exceed these limits. Of our selected 500 pc candidates, 215 sources (37%) have RUWE and AEN assignments far larger than these criteria as shown in Fig. 1. Binary systems, in which many hot subdwarfs reside, may induce a photocentric wobble that is problematic for *Gaia*'s single-star astrometry model, resulting in an inflation of these values (Lindegren et al. 2018; Belokurov et al. 2020; Penoyre et al. 2022; Lindegren et al. 2021). Of the identified hot subdwarf stars in our sample, those hosting main sequence companions appear to facilitate the largest RUWE values up to 5.55, which is a tangible outcome as both constituents are of similar brightness. WD companions seem to have little impact by comparison.

For the cleaning of our catalogue, we adopt a relaxed limit of RUWE $< 7$ as indicated by the red dashed line which excludes 205 sources from our sample. This limit was chosen as it does not remove any spectrosicpcally confirmed hot subdwarf stars and is in the interest of arriving at a criterion that defines a sample that is both as clean and complete as possible.

The impact of this selection is evident in Fig. 2, showcasing *Gaia* CMD's in the top panels and skyplots in the bottom panels. All candidates from CG22, selected using Eq. 1, are depicted. Those with RUWE values exceeding 7 are highlighted red. Notably, many of these sources cluster near the MS of the CMD between 0.3 and 0.4 BP-RP, as well as in a specific region in the Galactic mid-plane at a longitude of ≈ 120 degrees. If genuine, such behaviours are unexpected, as the impact of interstellar extinction is unlikely to play a significant role in this selection (refer to Sect. 6 for detailed discussion). A 500 pc sample should demonstrate a roughly homogeneous distribution across the sky, as confirmed by a statistical test of homogeneity (see Sect. 4.2). Twenty-one sources with RUWE > 7 are identified as MS stars based on spectra in the LAMOST DR8 database. It is highly probable that the majority of these spurious sources result from contamination in crowded fields. The application of our adopted criterion, RUWE < 7, effectively eliminates these dense star concentrations, yielding a clean and homogeneously distributed sample shown in the right panels of Fig. 2, which serves as the basis for this study.

A comparable pattern was observed in sources from a 500 pc sample initially sourced from Geier et al. (2019), which, in turn, utilised *Gaia*'s DR2 (Gaia Collaboration et al. 2018b). This initial sample comprised 635 candidate sources, with 70% necessitating spectroscopic identification. The ongoing spectroscopic follow-up campaign, conducted since 2019, yielded the majority of spectra utilised for spectral classification. The release of *Gaia*'s DR3, featuring an extended base line, provided updated astrometric data with fewer outliers. In Fig. 3, a *Gaia* CMD pair-plot showcases the original 500 pc sample of 635 hot subluminous star candidates. The pair-plot (illustrated with dotted lines) visually depicts the shift these sources experienced in this parameter space due to the updated parallaxes and apparent magnitudes of DR3, with the circle indicating the DR3 data. Notably, many sources near the MS rejection criterion of CG22 (depicted by the dashed line) underwent a significant vertical shift, with the red circles indicating those possessing RUWE values larger than 7. Through our extensive follow-up spectroscopic campaign, we determined that 295 (99%) of these removed sources are MS stars located in crowded fields near the Galactic plane. The revised astrometry from *Gaia's* DR3 now places all these sources well beyond 500 pc. In this paper, we consider this evidence sufficient to confidently establish a limit in RUWE for the cleaning of the 500 pc sample, particularly for those sources now sourced from *Gaia's* DR3. It is important to emphasise, however, that no single, uniform selection criterion can be universally applied across the sky due to the intrinsic limitations of *Gaia*, such as challenges in crowded fields, which cannot be encapsulated by a single threshold in all situations. We underscore that our selection criteria are specifically tailored for this sample, and any adopted quality control criteria should be chosen appropriately for each unique selection of stars from the *Gaia* database.

---
[1] https://simbad.u-strasbg.fr/
[2] Details on the RUWE can be found in the document "Renormalising the astrometric chi-square in *Gaia* DR2" and is available at: https://www.cosmos.esa.int/web/gaia/public-dpac-documents

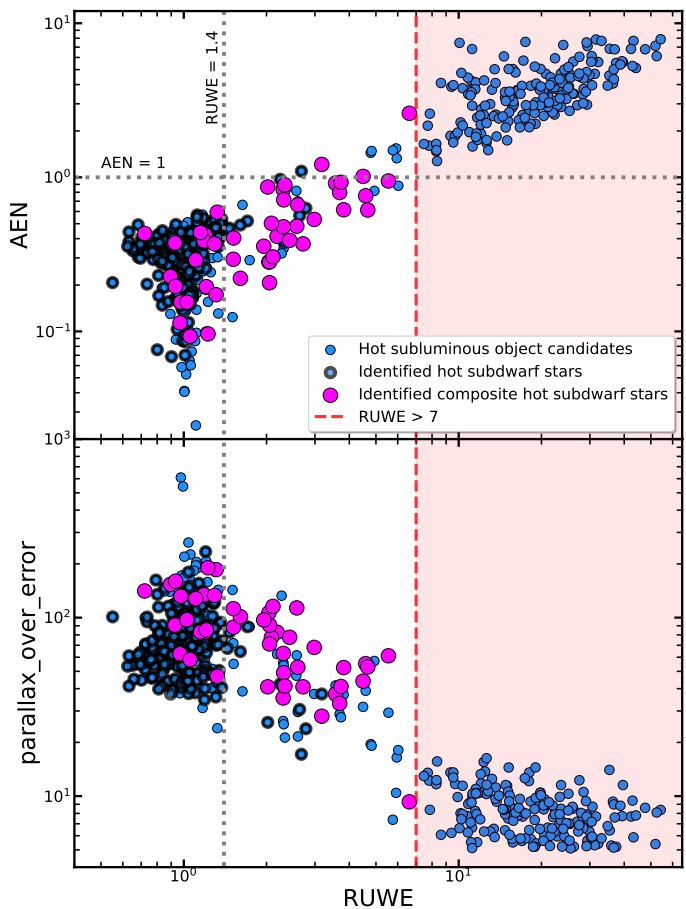

Fig. 1: Plot depicting the distribution of the 584 candidate hot subluminous stars within 500 pc (blue) for three astrometric quality control parameters provided by *Gaia* DR3. Top panel: AEN vs RUWE where all currently known hot subdwarf composite and non-composite binaries are given as magenta and black circles, respectively. The dashed grey horizontal and vertical lines indicate an AEN of 1.0 and a RUWE of 1.4, respectively. The red dashed line and the corresponding red shaded area represent the rejected sources in the sample. Bottom panel: Same as top, but with parallax_over_error vs RUWE.

Lastly, a handful of targets were also removed and were not subject to follow-up spectroscopic observation and are instead listed in Table A.3 and indicated as removed sources. These systems were either in close proximity to the spurious region of our data in Fig. 2 and still possessed high RUWE values, or were deemed too bright to be genuine hot subdwarf stars.

## 3. Observations, data reduction and classification

### 3.1. Spectroscopy

Over two-thirds of the *Gaia* DR2 500 pc sample were primarily based on colour and absolute magnitudes, triggering a spectroscopic follow-up campaign that observed over 400 hot subdwarf stars between 2019 and 2021. Observations were carried out from a variety of telescopes and instruments including the INT/IDS,[3] NOT/ALFOSC,[4] NTT/EFOSC2,[5] SOAR/Goodman,[6] and CAHA 3.5m/PMAS[7] and targeted the optical sector between 3300 Å and 7400 Å where a high signal-to-noise of at least 50 was obtained for each star which is suitable for classification and a detailed line profile analysis. A summary of the observations is presented in Table 2, which covers a total of 34 nights where some targets were observed more than once. In addition, high-quality spectral classifications for 322 of our target candidates were provided, which stem from a low-resolution survey conducted at the B&C spectrograph on the University of Arizona 2.3m Bok telescope. For a description of this dataset, see Green et al. (2008).

The spectroscopic data obtained by follow-up observations were reduced and analysed using PyRAF procedures (Science Software Branch at STScI 2012), a command language for the IRAF (Tody 1986) (Image Reduction and Analysis Facility) which is a general purpose software developed by the National Optical Astronomy Observatories (NOAO) using the longslit package for 2D spectroscopic images. This included basic bias and flat-field corrections, wavelength calibrations, and flux calibrations of the instrument response function with atmospheric extinction taken into account.

### 3.2. Archival data

Online databases provided the remaining spectroscopic data. The Large Sky Area Multi-Object Fiber Spectroscopic Telescope DR7 (v.2) (LAMOST; Luo et al. 2022) low-resolution spectroscopic survey (LRS), and its more recent DR8, provided 151 quality spectra covering wavelengths up to 9100 Å. The European Southern Observatory (ESO) archive[8] and the Mikulski Archive for Space Telescopes (MAST)[9] were also searched and yielded several high-resolution spectra for stars in our sample which will be utilised for accurate atmospheric parameters. None of our selected targets are found in the archives of the Sloan Digital Sky Survey (SDSS) because they are brighter than the bright magnitude limit for this survey.

### 3.3. Hot subluminous star classification

Our classification of hot subdwarf stars is based on the scheme outlined in Moehler et al. (1990b), which is updated and extended in CG22. Figure 4 displays example spectra illustrating this scheme. We emphasise that the classifications in this work are primarily based on visual inspection of available or obtained spectra, or otherwise reliable literature classifications.

---
[3] http://www.ing.iac.es
[4] http://www.not.iac.es/instruments/alfosc/
[5] https://www.eso.org/public/teles-instr/lasilla/ntt/efosc2/
[6] https://noirlab.edu/public/programs/ctio/soar-telescope/goodman/
[7] https://www.caha.es/CAHA/Instruments/PMAS/pmas.html
[8] http://archive.eso.org/eso/eso_archive_main.html
[9] https://archive.stsci.edu/

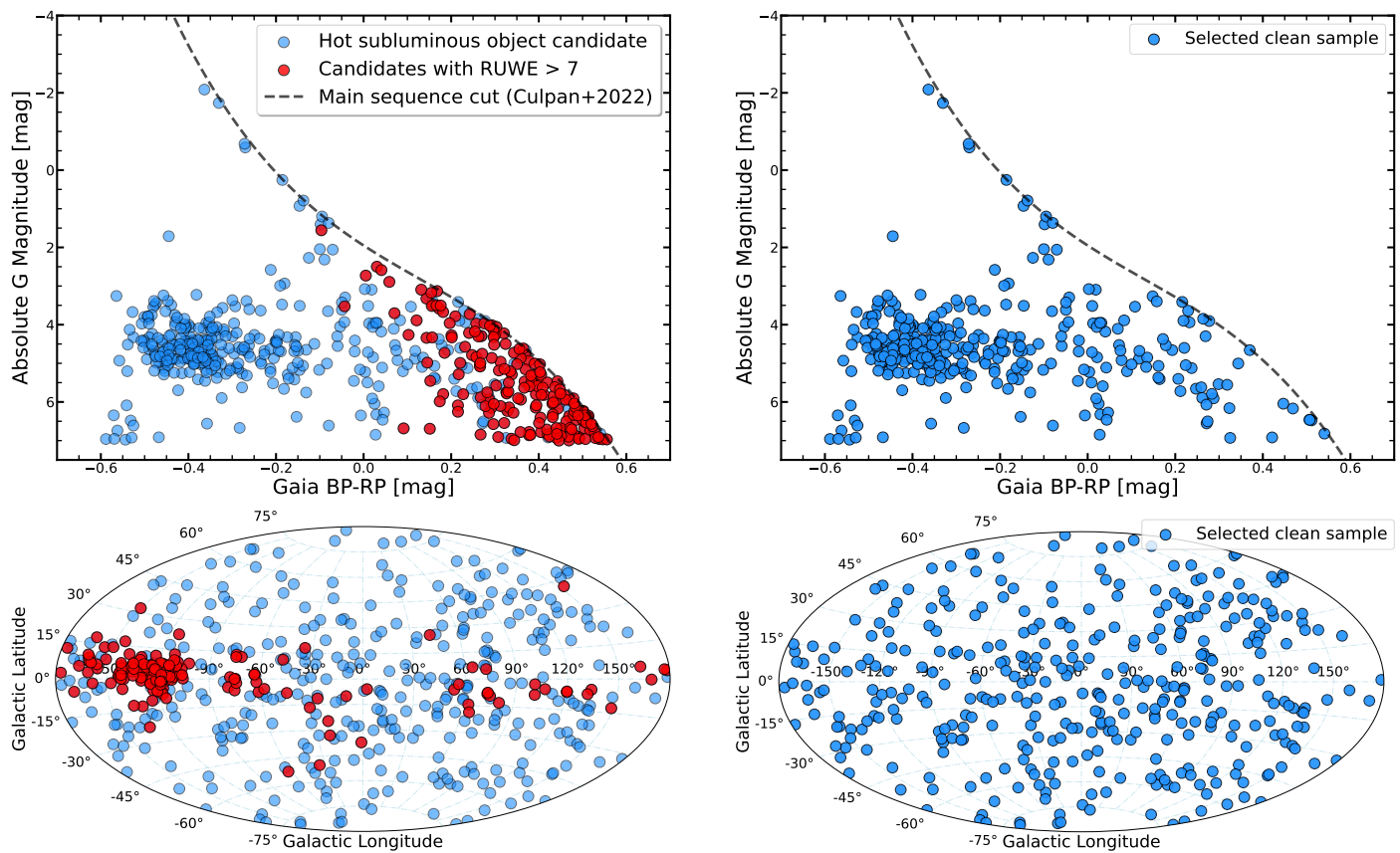

Fig. 2: *Gaia* colour-magnitude diagrams and sky distributions of the 584 candidate hot subluminous stars within 500 pc drawn from *Gaia* DR3. Those candidates with RUWE values in excess of 7 are indicated by red circles and have been removed in the panels on the right, which display the cleaned sample that was adopted. The exclusion of sources with RUWE>7 resulted in a reasonably homogeneous sky distribution given in the bottom right panel as expected for a 500 pc sample.

| Telescope/Instrument | Stars observed | Set-up | Wavelength range [Å] | Spectral resolution ($\Delta\lambda$ [Å]) | Programme ID |
|---|---|---|---|---|---|
| INT/IDS/EEV10 | 140 | R400V | 3300 - 7400 | 4.2 | ING.NL.19B.005 |
|  |  |  |  |  | ING.NL.20A.003 |
| NOT/ALFOSC | 13 | Grism 18 | 3450 - 5350 | 4.4 | 60-503 |
| NTT/EFOSC2 | 173 | Grism 7 | 3270 - 5240 | 7.4 | 0103.D-0511(A) |
|  |  |  |  |  | 0104.D-0514(A) |
|  |  |  |  |  | 0105.D-0121(A) |
|  |  |  |  |  | 0106.D-0188(A) |
|  |  |  |  |  | 0103.D-0530(A) |
| SOAR/GOODMAN | 29 | 400 1 mm$^{-1}$ | 3300 - 7050 | 2.97 | 2019B-012 |
|  |  |  |  |  | 2021A-008 |
|  |  |  |  |  | 2022B-966788 |
|  |  |  |  |  | C057 |
|  |  |  |  |  | CN2020A-87 |
|  |  |  |  |  | CN2020B-74 |
|  |  |  |  |  | CN2021A-52 |
| CAHA 3.5m/PMAS | 11 | V600 | 3600 - 6800 | 3.63 | GTO |
| Other |  |  |  |  | Source |
| LAMOST DR8 | 151 | LRS | 3690 - 9200 | 3.05 | Database query |
| Literature parameters | 149 |  |  |  | Culpan et al. (2022) |

Table 2: Overview of the spectroscopic data

In addition to this scheme, hot subdwarf stars may display signatures of cool companions in the spectrum. We take the *G*-band absorption as a clear companion signature, as well as the Mg I triplet at 5167 Å, 5173 Å, and 5184 Å. The photometric signatures of composite-colour hot subdwarf stars are described in detail in Sect. 5.3 where the construction of spectral energy distributions are employed to detect cool MS companions. This is necessary since companions of type K may be too dim to be seen in an optical spectrum, whereby the SED method then becomes

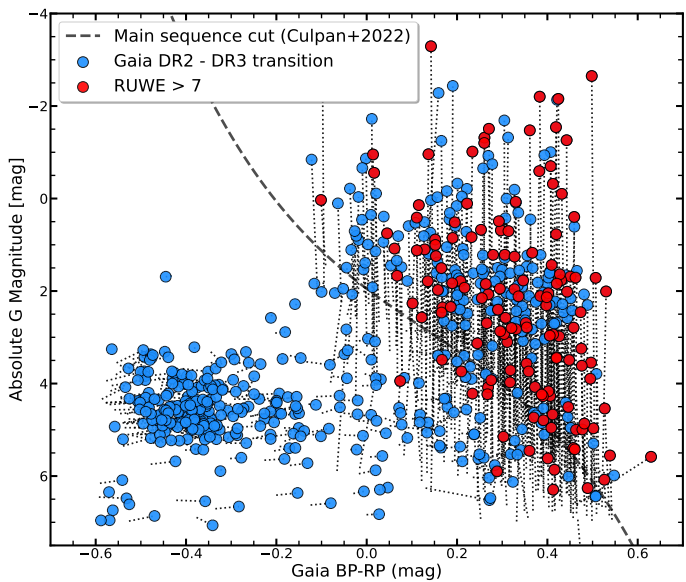

Fig. 3: Colour-magnitude diagram of hot subluminous star candidates originally selected from Geier et al. (2019), which draws from *Gaia* DR2. The plot highlights the impact the updated parameters in the *Gaia* DR2 to *Gaia* DR3 transition had on the flagship 500 pc sample. The pair-plot shows the shift each star underwent after DR3 where the circle represents the DR3 data. The vertical shift near the MS stems primarily from the improved astrometric solution.

the only option to detect the companion in the infrared if the photometric data is available.

As was introduced in CG22 to compliment the spectroscopic classes sdB, sdOB, sdO, and He-sdB/OB/O, we also consider the spectral classes O(H), O(He), PG1159, and [WR] to classify the hotter central stars of planetary nebulae (CSPNe) and other post-AGB stars found in our parameter space. In addition, this region in the HRD has been found to also contain cooling progenitors of helium WDs such as HD 188112 (Heber et al. 2003). These stars may be spectroscopically indistinguishable from sdB stars, as is the case for HD 188112. However, they may also display spectral features similar to MS A-type stars that is void of any helium such as [PS72]97, a recently identified helium WD progenitor that is present in our sample (Kosakowski et al. 2023). We use the sdA class to distinguish this population.

Lastly, we use the spectral classes of WD, CV, MSB/A/F, and BHB in order to fully classify our sample where no known spectroscopic classes are otherwise available in the literature. The last two are often indistinguishable using low-resolution spectra. For this population, we combine our obtained spectra with SED fits (Sect. 5.3) to obtain the radius and approximate masses of these stars to distinguish them.

This scheme is adopted in an attempt to comprehensively and homogeneously account for all the populations found in this parameter space. A detailed analysis of the subpopulations of hot subdwarf stars will be explored in a future paper to disentangle the various evolutionary histories.

## 4. Analysis

### 4.1. Spatial distribution, space density and scale height

Following the spectral classification by visual inspection, we explore potential relationships between the classes of hot subluminous stars and their Galactic positions. Previous studies faced challenges from complicated selection biases in magnitude-limited surveys or were constrained to specific sky regions, especially for hot subdwarfs (e.g. Heber 1986; Downes 1986; Villeneuve et al. 1995a,b) (see Table 1 for an overview). The sky plots on the left in Fig. 5 reveal a relatively homogeneous distribution of hot subdwarf stars (top left panel) and other constituents like CVs (bottom left panel) across the sky, irrespective of spectral class. On the right, we project the sample onto the Galactic Z-R and Y-R planes; the hot subdwarfs again exhibit a homogeneous distribution in 3D space, with a small void of stars in the upper right quadrant, possibly indicative of Galactic structure (see Sect. 4.2). A tendency for stars to concentrate near the Galactic plane suggests a decreasing space density with Galactic height. CVs, in contrast, show indications of scarcity towards the Galactic centre, while WDs are nearly absent on one hemisphere. These systems, primarily located in the lower right corner of the CMD (see Fig. 11 in Sect. 5), are more likely to be removed from our selection due to interstellar extinction.

It is often assumed that the vertical density profile of the Galaxy at the solar radius can be described in the same way as the radial profile, a simple exponential barometric distribution for various kinds of stars (e.g. Kroupa 1992; Jurić et al. 2008; Bovy et al. 2012, 2016; Pala et al. 2020; Schaefer 2022), which would then take the form:

$$\rho = \rho_0 \, e^{-|z|/h_z} \qquad (2)$$

where $h_z$ is the exponential scale height, $z$ is the distance in parsecs from the Galactic mid-plane, $\rho$ is the density at distance $z$, and $\rho_0$ is the density at $z = 0$. However, there are currently two main geometric contenders of the $z$-profile, the second being a hyperbolic secant, which takes the following form:

$$\rho = \rho_0 \, \text{sech}^2 \left( \frac{|z|}{h_0} \right) \qquad (3)$$

where $h_0$ is the vertical characteristic height of a self-gravitating, locally isothermal sheet (Spitzer 1942; van der Kruit & Searle 1981a,b). This profile essentially combines a Gaussian at small distances around the mid-plane with a barometric exponential at large distances and is in popular use (e.g. Gilmore & Reid 1983; Villeneuve et al. 1995a,b; Bilir et al. 2006; Yoachim & Dalcanton 2006; Widrow et al. 2012; Ma et al. 2017; Canbay et al. 2023). To facilitate comparison with other literature values, we take the approximation of this function at $|z|/h_0 \gg 1$ where it asymptotes to resemble an exponential with a scale height $h_z = h_0/2$. While other forms of this model, such as *sech*, exist, we do not explore them here due to their minimal impact on describing the plane's flattening with the relatively small number of stars in our sample. Multiple components are often employed to acquire more precise descriptions of the data, especially to account for multiple Galactic populations. Given the reported scale heights of the Galactic thin and thick disk (300 pc and 900 pc, respectively, according to Jurić et al. (2008)), we adopt a straightforward approach, mapping the $z$-axis density variation using a single component. We use both a hyperbolic secant (eq. 3) and an exponential (eq. 2) as two-parameter fits for $\rho_0$ and $h_z$ and compare the results. The Galactic scale length is not considered here, as the density variation along the Galactic $x$-axis is less than 25% on the scale of 1 kpc at the solar radius (Binney & Tremaine 1987), and very few of our stars are located at the extremities of our sample as projected onto the $x$-axis.

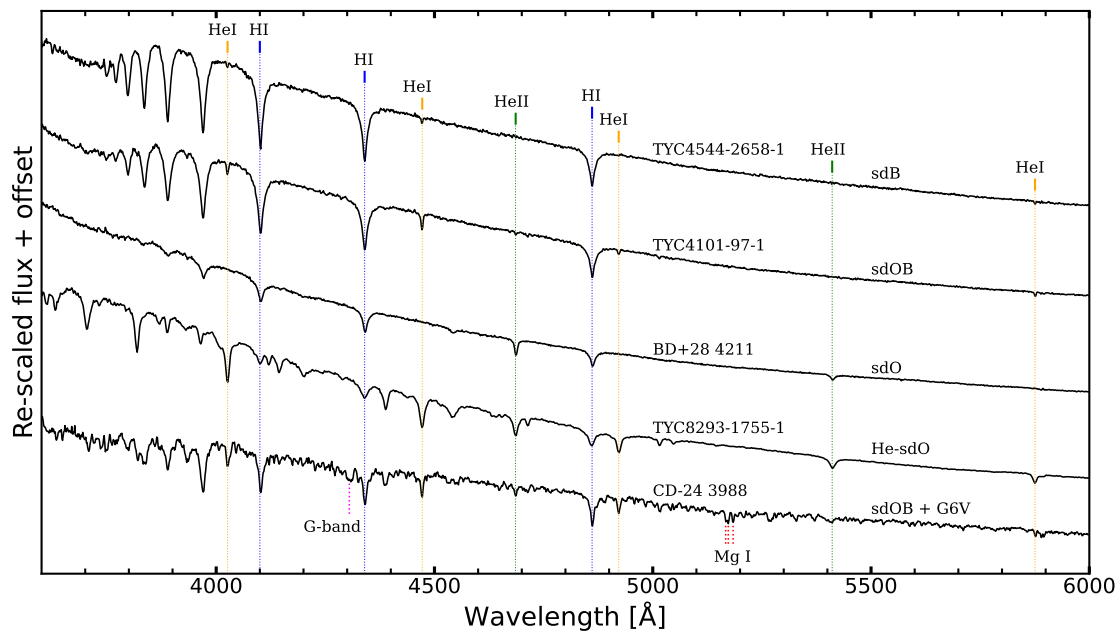

Fig. 4: Example identification spectra obtained in this study to illustrate the adopted classification scheme and to showcase the high quality of the dataset. The primary diagnostic lines of He I, He II, H I, the Mg I triplet, and the G-band are marked in gold, green, blue, red, and magenta respectively. These spectra were obtained at the INT and at SOAR (see Table 2).

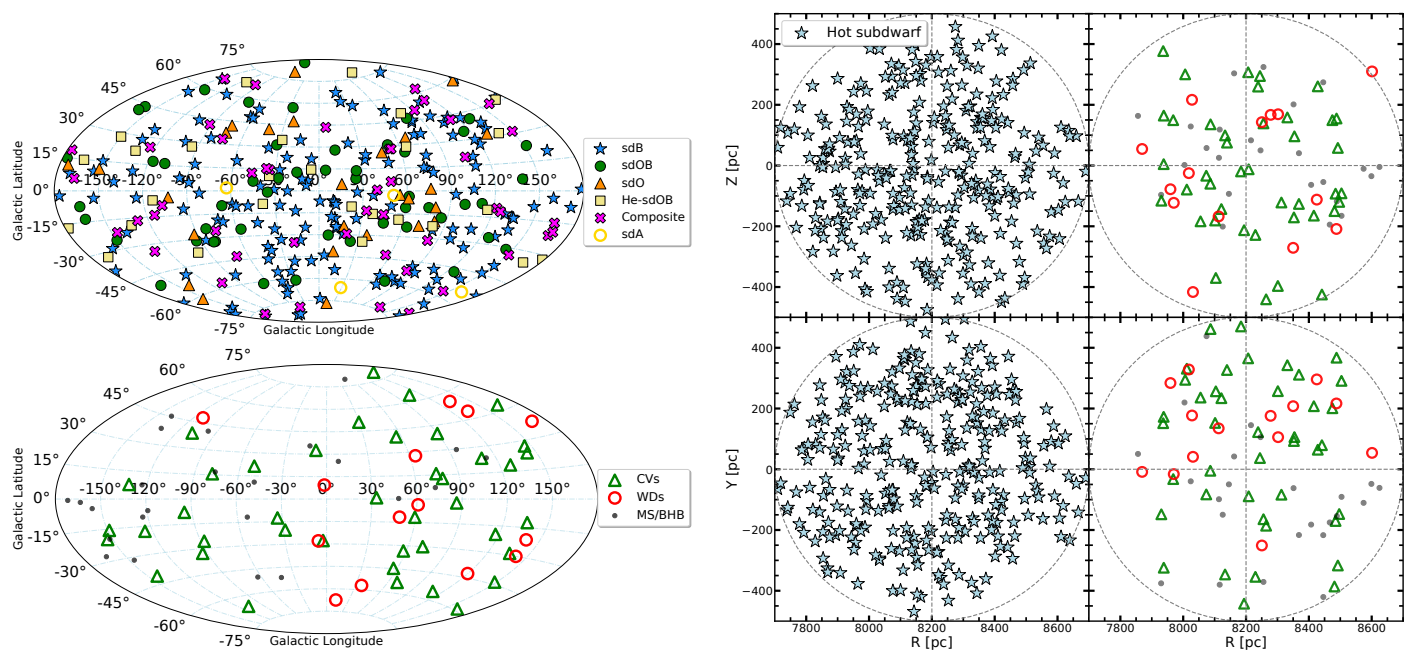

Fig. 5: Galactic distribution of the 500 pc sample and its constituents, which are labelled accordingly for all plots. The left panel describes the projected distribution of the sample on the sky in terms of Galactic longitude and latitude, whereas the right panel gives the projected positions onto the Galactic plane ($X-Y$ plane) and on the plane perpendicular to it ($X-Z$ plane). Here the $X$-coordinate is given in terms of $R$, the solar distance from the Galactic centre.

We focus initially on the non-composite hot subdwarfs, constituting the primary population in our sample with 257 members (table A.1). Composite-colour hot subdwarfs are excluded from this analysis due to known incompleteness, which could introduce bias (see Sect. 6 for a discussion on the completeness of this population). The sample is divided into equally spaced bins along the $z$-axis. To increase number statistics, we consider the absolute value of the *z*-component of each star, essentially doubling our sample, as illustrated in Fig. 6. The component *z*, denoted as $z = d\sin(b)$, incorporates the Galactic latitude *b* and the distance *d* of each star. The solar offset to the Galactic plane is not considered in this analysis due to its omission from the selection criteria, and its determination is not confident with the limited stars in our sample. We generate arrays of distances using scalar parallax measurements from *Gaia*, along with the inflated parallax uncertainties (El-Badry et al. 2021), employing a Monte Carlo approach. These distances are projected onto the *z*-axis, forming distance probability distributions (see the right panel of Fig. 6 for a stacked histogram of these distributions). The smoothing in the distribution, reflecting increasing parallax errors with distance, is evident. This procedure accounts for the non-linearity of parallax inversion, especially near the sample boundary or between bins, where stars contribute with varying weights due to distance uncertainties. The probability arrays for each bin provide the number of stars and uncertainties, determined by means and standard deviations, respectively. Calculating densities for each bin involves dividing these numbers by the corresponding volumes of each slice, $V_{slice}$, determined using the equation:

$$V_{slice} = \frac{1}{6}\pi h(3R_1^2 + 3R_2^2 + h) \quad (4)$$

where *h* is the width of the spherical segment or bin width and $R_1$ and $R_2$ are the radii of the top and base of the circle segment.

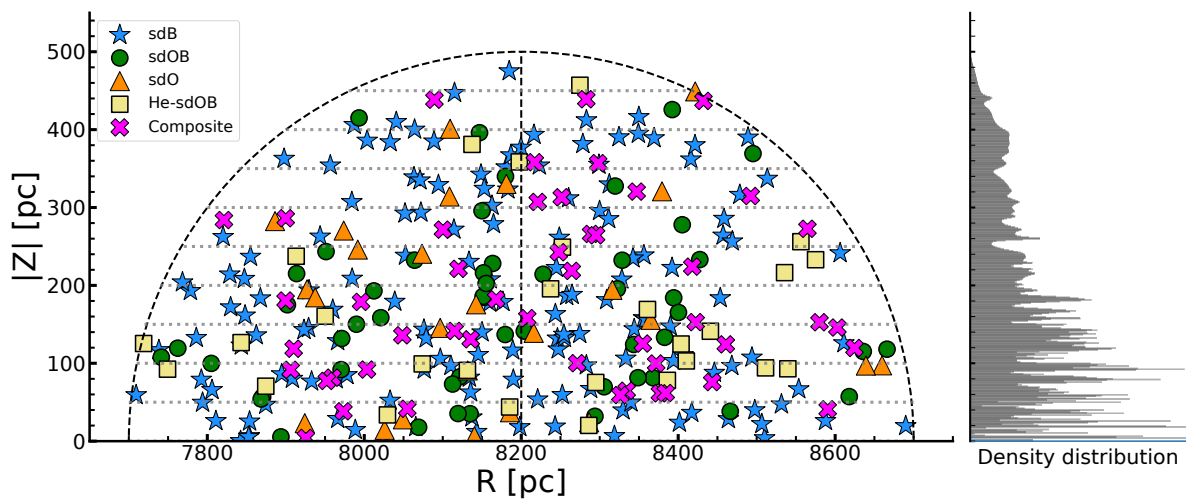

Fig. 6: Projected distribution of the 500 pc sample above the Galactic plane. Left panel: 500 pc sample is projected onto the R-|Z| plane with the different classes of hot subdwarfs colour-coded accordingly. The black dashed line indicates the 500 pc distance limit, whereas the horizontal grey dashed lines represent a 50 pc binning example for how the sample has been divided up to derive the scale height. Right panel: Positional probability density distribution projected onto the z-axis taking the standard parallax errors of *Gaia* into account - displayed as a stacked histogram.

We employed a non-linear least-squares algorithm (Python library LMFIT) to simultaneously fit the scale height and mid-plane density, as given by equations 2 and 3. Errors were obtained from the estimated covariance matrix during the fit. Our analysis indicates that the $\text{sech}^2$ provides a much better description of the data, especially around the Galactic mid-plane. An exponential fit, on the other hand, performs well when modelling beyond 100 pc along the *z*-axis. However, attempting to model all data points results in a systematic inclination for the fitting procedure to choose excessively large scale heights ($h_z \geq 1000$ pc). This reflects a smooth profile at $z \leq 100$ pc. The implications of this tendency are unclear; it could, although unlikely, signify the true physical morphology of the local thin disk or imply a significant number of missing stars in the mid-plane, warranting further investigation.

The number count versus *z* is shown in the top panel of Fig. 7 for the median bin sizes of the fitting routine described below, with the corresponding density profile provided in the lower panel. A notable feature in this plot is a peak between 300 and 400 pc, which probably arises from the rapidly decreasing bin volumes along the along the *z*-axis.

However, vertical wave-like features in the number counts that run parallel to the Galactic disk have been corroborated in several studies (Widrow et al. 2012; Yanny & Gardner 2013; Ferguson et al. 2017; Xiang et al. 2018; Bennett & Bovy 2019). Widrow et al. (2012) found peaks in the stellar number density at around $Z \approx -400$ and 800 pc for the solar neighbourhood, which are asymmetrically reflected around the Galactic plane, and is reportedly strongest for the youngest populations in the more recent work by Xiang et al. (2018). These findings are in reasonable agreement with our local population of hot subdwarf stars, particularly the peak at $Z \approx -400$ pc (see Fig. 8).

Due to the number fluctuations, the choice of bin size for our *z*-profile fits introduces a relative uncertainty on the resulting scale height of up to 25% and up to 10% on the mid-plane density. To address this, we perform two parameter fits on the scale height and mid-plane density using a hyperbolic secant profile for all integer divisors of 500 between and including 5 and 20, which corresponds to bin widths between and including 100 pc and 25 pc, respectively. Attempting to fit bin widths outside of this range results in either empty bins with large fluctuations or too few data points for the fit, respectively. Along with the statistical errors from each fit propagated through from the estimate covariance matrix, we also incorporate Poisson uncertainties in quadrature. This approach accounts for the parallax uncertainties of the individual objects in our sample and the statistical variations due to the limited sample size. Poisson errors for the mid-plane densities are calculated with the following:

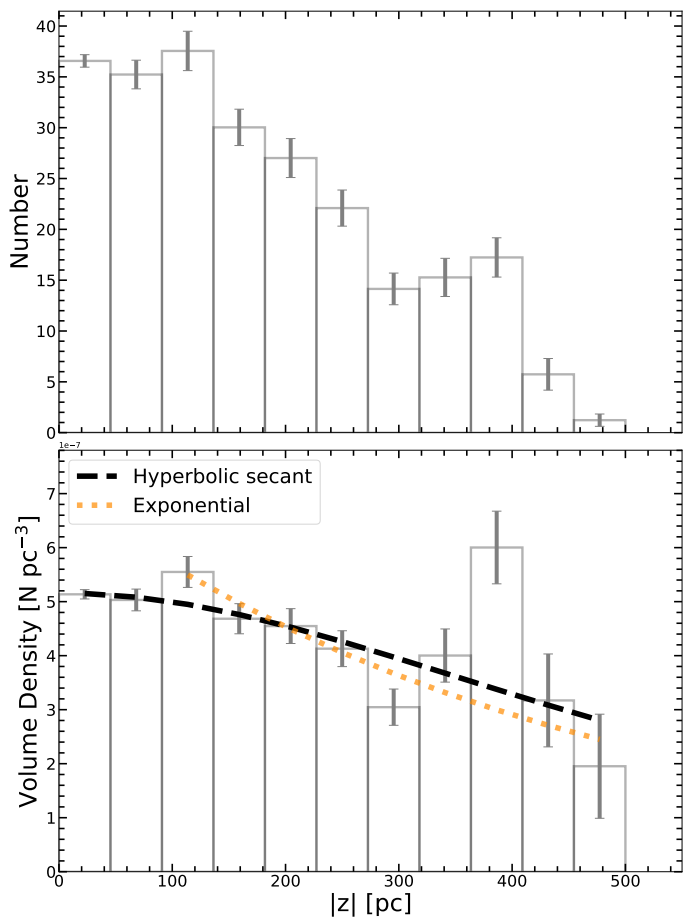

Fig. 7: Distribution of the non-composite hot subdwarf stars over the perpendicular distance to the Galactic plane, |z|. The top panel gives the number count, whereas the bottom panel displays the volume density. Median bin sizes of the fitting routine are displayed. Example fits of a hyperbolic secant profile (black) and an exponential (gold) are shown; the latter has only been fit for |z| > 100 pc. The error bars are generated from the standard deviations of the positional probability distributions.

$$\sigma_{Poisson} = \frac{\sqrt{N}}{V_{\text{eff}}} \quad (5)$$

where $N$ is the number of stars in our sample and $V_{\text{eff}}$ is the effective volume of our sample given by Eq. 7. Poisson errors are not calculated for the scale height, as we assume the distributions of each possible sample to be the same. By fitting all possible equally spaced bins, we find an average scale height and mid-plane density of $h_z = 281 \pm 62$ pc and $\rho_0 = 5.17 \pm 0.33 \times 10^{-7}$ stars/pc$^3$, respectively, for the non-composite hot subdwarf stars in our 500 pc sample. This suggests a primarily younger thin-disk population, with non-composite hot subdwarfs making up $84.1^{+2.3}_{-1.9}\%$ of the sample, assuming binomial uncertainties. To ensure that this is representative of the full population of hot subdwarfs in our sample, we then incorporate all known composite-colour hot subdwarfs within 500 pc. This inclusion raises the mid-plane density to $\rho_0 = 6.15^{+0.55}_{-0.53} \times 10^{-7}$ stars/pc$^3$. The errors in this density calculation are determined by propagating standard errors on parallax measurements, Poisson uncertainties (as given by Eq. 5), and binomial uncertainties on the fraction of non-composite hot subdwarf stars in our sample. We assume that the composites share the same characteristics as the non-composites and are distributed similarly in 3D space, maintaining the scale height. The final mid-plane density calculation includes an asymmetric upper error bar, accounting for the estimated number of missing composite hot subdwarfs in our sample (as discussed in Sect. 6). Consequently, the revised mid-plane density is $\rho_0 = 6.15^{+1.16}_{-0.53} \times 10^{-7}$ stars/pc$^3$.

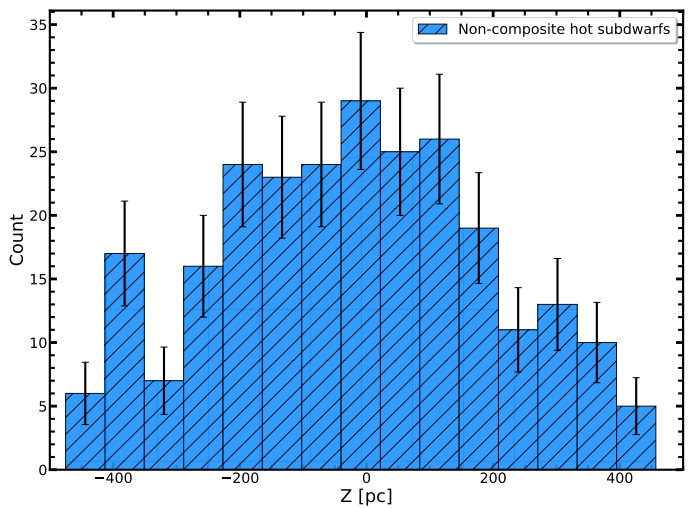

Fig. 8: Number histogram of z-distances for the non-composite hot subdwarfs around the Galactic plane.

### 4.2. A test of a uniform distribution

To evaluate the statistical significance of the local density fluctuations observed in our sample, we employ the Kolmogorov-Smirnov (KS) test to assess its homogeneity. Additionally, this test helps discern whether the scale height is the primary factor contributing to the decline in density along the z-axis. In their fifth catalogue of nearby stars, Golovin et al. (2023) utilised the KS test to evaluate the completeness of their 25 pc sample by comparing the empirical Cumulative Distribution Function (CDF) to an analytical CDF given by:

$$\text{CDF}(r) = \frac{r^3}{R_0^3}, \quad 0 \le r \le R_0 \quad (6)$$

where $R_0$ is for a sphere of radius 25 pc in their case. A deviation from homogeneity implied the completeness limit at that magnitude. As our 500 pc stars are relatively bright, our sample does not suffer this magnitude limitation. We followed a similar approach and compared the CDF of the full sample of hot subluminous stars to the CDF of that given by Eq. 6 for $R_0 = 500$ pc and performed a KS test over 5 pc intervals to test for homogeneity. Attempting to use subsamples in this analysis, such as the confirmed non-composite hot subdwarf stars as was done in Sect. 4.1, results in unreliable measurements as low-number statistics begins to dominate. The results of this effort are given in Fig. 9 as a blue line. A high p-value for distances between 100 and around 350 pc is seen, beyond which it begins to drop and suggests a deviation from homogeneity which is in reasonable agreement with our determined scale height. This becomes statistically significant at a distance of 440 pc where it reaches the $p = 0.05$ threshold. We now follow the prescription of Inight et al. (2021) and define the effective volume $V_{\text{eff}}$:

$$V_{\text{eff}} = \int_0^R \int_{-\sqrt{R^2-x^2}}^{\sqrt{R^2-x^2}} (2\pi x) \operatorname{sech}^2\left(\frac{|z|}{2h_z}\right) dz\, dx \qquad (7)$$

where $x$ points in the direction of the Galactic centre, $z$ is perpendicular to the Galactic plane, and $h_z$ is the scale height along the $z$-axis. For a decreasing scale height, Eq. 7 yields an increasing volume with respect to a simple spherical volume define by $4\pi R^3/3$. Using Eq. 7, we can correct for the scale height along the $z$-axis by adopting our determined value of 281 pc for $h_z$. In doing so, the same KS test then results in the red line displayed in Fig. 9. This time, the empirical CDF of the sample appears to remain homogeneous at all distances and is statistically indistinguishable from the analytical CDF at the 5% level. The slight dip present at around 450 pc could reflect an under- or overdensity at this distance or specifically the small void seen in Fig. 5 in the upper right quadrant in the Z-R and Y-R planes. However, we must be careful when interpreting this result, as the sample selection in CG22 is not straightforward and the inclusion of other low-luminosity stars such as CVs and WDs may bias the result. Yet, as the sample appears to be homogeneous overall, this rules out any major systematic selection effects or biases for our clean selection of stars and is a striking result given the relatively few stars in our sample spread across a large volume. Moreover, it reinforces our determined scale height as being the dominant contributor to the variation in spatial distribution across the 500 pc sample where the local under- and overdensities can be interpreted as statistically insignificant.

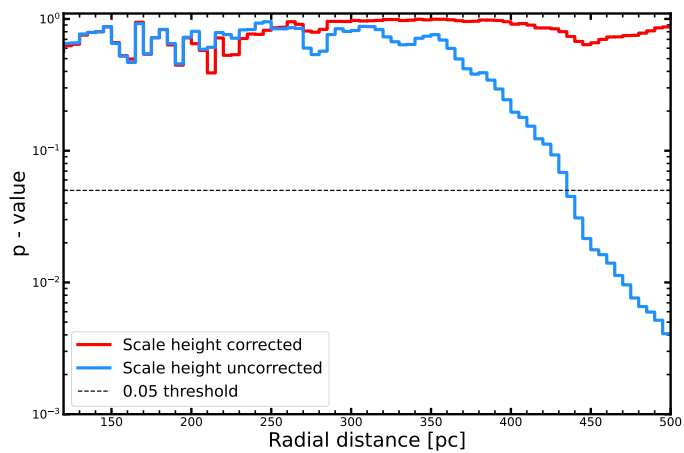

Fig. 9: Plot displaying the determined p-value as a function of radial distance at 5 pc intervals from the KS test. The blue line gives the first iteration and displays a deviation from homogeneity around 350 pc, whereas the red line gives the second iteration when the scale height is taken into account and remains high, suggesting homogeneity.

Finally, in addition to these tests over distance, we use the KS test to probe the overall homogeneity of the non-composite hot subdwarfs in the sample. We generate 1000 mock samples when taking the scale height into account where has a distribution of distances according to:

$$r = R_0\, y^{a/3} \qquad (8)$$

where $y$ is a random number between 0 and 1, and $a$ describes a correction factor for the scale height. Each sample thus represents a distribution of distances for which we can randomly draw a number of objects from to compare to the empirical CDF of our non-composite hot subdwarfs. We then randomly draw 257 data points from each mock sample and perform the KS test, which yields an average p-value of 0.3. This implies that this population is statistically homogeneous overall and that no major selection bias is in effect.

### 4.3. Kinematics

A measurement of the scale height provides insights into the age of a stellar or substellar population, with a steeper scale height generally yielding a higher mid-plane density, inferring a younger population. However, most hot subdwarfs lack age information due to strong diffusive effects in their atmospheres, rendering metallicity an unreliable age indicator. Instead, we can examine the kinematical properties of a stellar population, which is also linked to their age. Older stars tend to have larger kinematic dispersions than younger stars (Bovy 2017; Sanders & Das 2018). Utilising the accurate proper motion measurements from *Gaia*, we calculate tangential velocities, which serve as proxies for dynamical heating, using the following formula:

$$V_T = \frac{4.74}{\omega_{\text{zp}}} \times \sqrt{\mu_{\text{RA}}^2 + \mu_{\text{DEC}}^2} \qquad (9)$$

where $\omega_{\text{zp}}$ is the zero-point corrected parallax, $\mu_{\text{RA}}$ and $\mu_{\text{DEC}}$ are the *Gaia* proper motion measurements in right ascension and declination respectively, and 4.74 is a unit conversion factor to give the tangential velocity $V_T$ in km/s. Figure 10 displays the distribution of tangential velocities as a histogram, where confirmed hot subdwarfs are shown in blue, and the remaining objects in the sample are grey. Both sets exhibit a similar distribution, indicating the absence of outstanding contamination arising from highly underestimated distance measurements in the Galactic plane, which was observed in DR2. Four stars in the sample have tangential velocities exceeding 200 km/s, as determined from *Gaia's* proper motion measurements. These include three known hot subdwarf stars: LS IV −14 116 (He-sdOB), Feige 46 (He-sdOB), and SB744 (sdOB+F/G). The fourth star, TYC144-2049-1, is a BHB star candidate according to Culpan et al. (2021). It is not surprising that a sample extending to 500 pc above and below the Galactic plane may include a significant fraction of kinematically heated sources, suggesting older thick-disk or halo population stars. Two of the four stars with tangential velocities exceeding 200 km/s, Feige 46 and LS IV −14 116, confirmed as halo stars (Dorsch et al. 2020), exhibit no RV variability. SB744 is also identified as a halo star with no RV variability (Németh et al. 2021). While it is unclear whether TYC144-2049-1 is a MS or BHB star, its measured RV of 211 ± 6 km/s, suggests it is probably a member of the halo population. These stars are not excluded in the scale height and space density calculations as it has marginal impact on the results. Obtaining reliable systemic velocities of all members of this sample for a kinematic analysis is the ambition of a future paper.

## 5. Content of the catalogue

In this section we detail the content of the 500 pc sample in terms of the variety of astrophysical objects found in the *Gaia* CMD

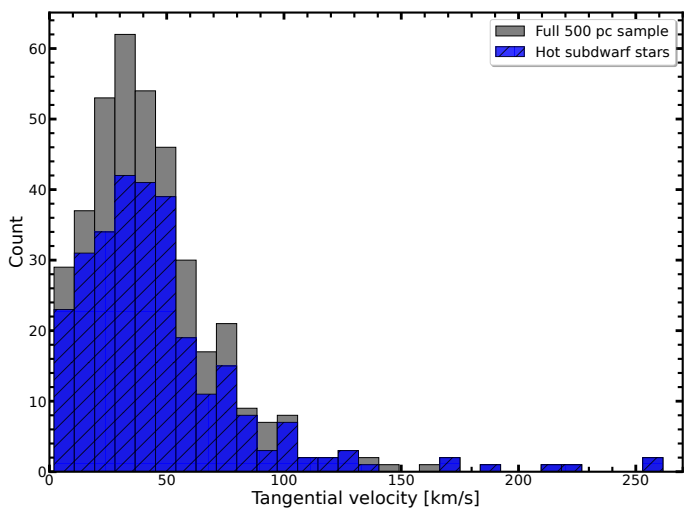

Fig. 10: Histogram displaying the tangential velocity of the full 500 pc sample (grey) and the confirmed hot subdwarf stars (blue) calculated using *Gaia's* proper motion measurements.

parameter space as illustrated in Fig. 11, which is accompanied by table 3.

| Spectroscopic class | Number | Fraction |
|---|---|---|
| Hot Subdwarfs (table A.1) | 305 | 77% |
| sdB | 178 | $58.4^{+2.9}_{-2.8}\%$ |
| sdOB | 65 | $21.3^{+2.2}_{-2.5}\%$ |
| sdO | 32 | $10.5^{+1.5}_{-2.0}\%$ |
| He-sdB/OB/O | 30 | $9.8^{+1.5}_{-2.0}\%$ |
| sdO/B+MS (table A.2) | 48 | $15.7^{+1.9}_{-2.3}\%$ |
| Other constituents (table A.3) | 92 | 23% |
| CV | 41 | $44.6^{+5.0}_{-5.2}\%$ |
| WD | 11 | $12.0^{+2.6}_{-4.2}\%$ |
| WD+MS | 3 | $3.3^{+1.0}_{-3.0}\%$ |
| sdA | 4 | $4.3^{+1.3}_{-3.2}\%$ |
| MS/BHB | 26 | $28.3^{+4.2}_{-5.1}\%$ |
| Removed | 7 | $7.6^{+1.9}_{-3.7}\%$ |
| **Overall total** | **397** | 100% |

Table 3: Catalogue classification.

### 5.1. Hot subdwarf stars

The sample is primarily made up of hot subdwarf stars (77%) which are listed in table A.1 and table A.2 for the non-composites and composites, respectively, and reports 83 new discoveries. Most previous dedicated surveys of hot subluminous stars have avoided the crowded and dusty regions of the Galactic mid-plane and were thus restricted to high Galactic latitudes (Vennes et al. 2011). It comes as no surprise then that most of the newly discovered hot subdwarfs in this programme are concentrated in the disk. This is best illustrated in Fig. 12, where a clear deficit of objects near the Galactic plane is seen in the previous catalogue of known hot subdwarf stars. The number of missing non-composite hot subdwarf stars due to interstellar extinction in this 500 pc selection is expected to be minimal (see Sect. 6 for an estimate of the completeness). This work builds upon previous studies and finds the majority of new objects below $b \approx 30$ degrees. We further report that 12 of the new systems show an infrared excess in their spectral energy distributions (Sect. 5.3) revealing them as possible composite-colour binaries. Four systems that were previously known, but were thought to be single hot subdwarfs, also show an infrared excess and are newly identified composite-colour binaries. HD 76431 is known to be a post-HB star (Khalack et al. 2014) which appears to be evolving through our selected parameter space and is positioned above the hot subdwarf cloud in the CMD (see Fig. 11, where HD 76431 is the only sdB with G < 2). With a spectrum resembling an sdB, it will be counted as such in this work.

Compared to the most recent previous catalogues of hot subdwarf stars (specifically Geier et al. 2017; Geier 2020; Culpan et al. 2022), we find the fractions of non-composite sdB and sdOB stars ($50.2^{+2.9}_{-2.2}\%$ and $16.7^{+1.9}_{-2.4}\%$, respectively) to be somewhat different, yet the total number of sdB + sdOB stars is exactly the same (~ 67%); this is probably due to the use of low-dispersion (published) spectra, where the weak He II 4686 Å line, which would differentiate between sdB and sdOB, might be present but undetected.

### 5.2. Known hot subdwarf stars in close binaries

Dedicated radial velocity variability (RVV) searches have indicated that at least one-third of hot subdwarfs reside in a close binary system with either a compact companion such as a WD, or a cool MS (dM) or brown dwarf (BD) (Maxted et al. 2001; Morales-Rueda et al. 2003; Napiwotzki et al. 2004; Copperwheat et al. 2011; Kawka et al. 2015; Kupfer et al. 2015; Geier et al. 2015, 2022). Currently, there are over 300 systems that host companions which are too faint to contribute any detectable flux to the observed light, yet are revealed by the Doppler shifts they induce in the spectral lines of the far brighter hot subdwarf. Additionally, they may also be revealed by the photometric variability they bring about in their light curves (e.g. Schaffenroth et al. 2019). Schaffenroth et al. (2022, 2023) carried out the most recent investigation on the nature of the close companions to hot subdwarf stars of type B (sdB) using data from the Transiting Exoplanet Survey Satellite (TESS) and the K2 space mission. Combining all the observational efforts listed above, a total of 36 systems, which either host WD, dM, or BD companions, are members of our 500 pc sample. This gives a close binary fraction of $12.3^{+1.7}_{-2.2}\%$ assuming a binomial distribution and without taking inclination into account. This fraction is probably a lower limit due to the 83 newly discovered systems which are yet to be investigated for signatures of binarity. This fraction has no impact on the analysis performed in this work as outlined in Sect. 4 as the close binary population is assumed to be complete (see Sect. 6). As such, an estimate of the fraction of this complete population could be directly used to constrain formation channels. An analysis of these systems, which will be the focus of a future paper, is vital for resolving the poorly understood short-lived phase of common-envelope evolution.

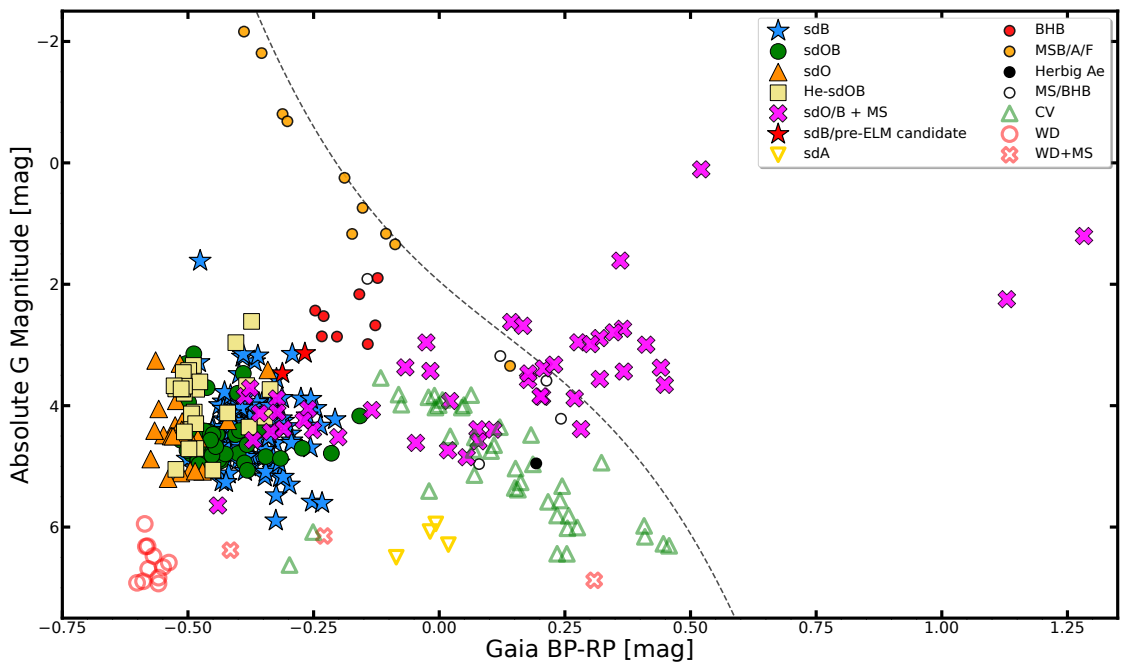

Fig. 11: Colour-magnitude diagram of all 397 members with clean astrometry (Sect. 2.2) in the 500 pc sample and colour-coded accordingly. Confirmed sources have been corrected for the effects of reddening (Sect. 6).

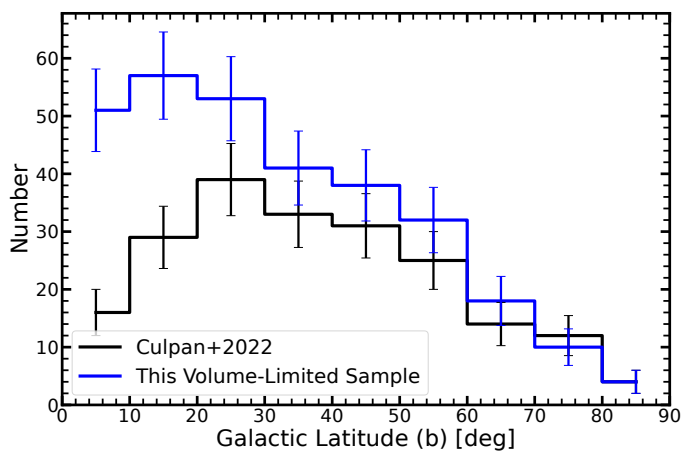

Fig. 12: Number distribution of the hot subdwarf stars in our volume-limited sample as a function of Galactic latitude, divided into nine bins of |b|. Black represents the catalogue of Geier (2020)/CG22 within 500 pc, whereas blue are those stars in this volume-limited edition. The error bars reflect statistical Poisson errors.

### 5.3. Hot subdwarf stars with cool main sequence companions

At least one-quarter of all hot subdwarfs are observed to host F/G/K-type MS companions (Girven et al. 2012; Solano et al. 2022) which are bright enough to be seen in the combined light of the system. However, the necessary cleaning during the construction of the catalogue of CG22 came at the expense of the exclusion of several known systems, in particular those hot subdwarfs located redwards of the empirical MS rejection criterion in the *Gaia* CMD. To date, there are 26 known sd+F/G/K systems with solved orbital parameters (Vos et al. 2012, 2013, 2017, 2019; Barlow et al. 2013; Otani et al. 2018; Deca et al. 2018; Molina et al. 2022; Otani et al. 2022). The study of such systems is crucial for constraining parameters of the stable RLOF channel (Vos et al. 2018; Molina et al. 2022) by which they are formed. It is essential to identify these systems as the population is known to be incomplete, and would otherwise bias the analysis of this paper. These companions are of MS types F, G, or K, which may dominate the optical spectral range and shift the systems CMD position into the MS. At least 18 such systems within 500 pc exist, including nine solved long-period systems, and are included in this catalogue. All manually added sources are indicated in table A.2 with an asterisk which mainly comprises those previously known composite binaries located within the MS, barring a single source that failed the `phot_bp_rp_excess_factor_corrected` criterion in CG22 (EC05053-2806), which probably stems from the bright F-type companion to the hot subdwarf. Many of these manually added systems host brighter companions of type F which we expect to be the least complete population.

To estimate the total number of composite systems we constructed spectral energy distributions (SEDs) for all candidates regardless of their RUWE values. These SEDs were used to identify hot subdwarf stars that exhibit atypically red infra-red (IR) colours, which can be attributed to the presence of a cool unresolved companion. Examples of a single and composite SEDs are shown in Fig. 13: K-type companions are detectable by their contribution in the IR, G-types contribute optical and dominate in the IR, and F-types tend to dominate in the optical as well; all of these companions are well detected by SEDs. In particular the K-type companions are well detectable in the SED, even if they are difficult to see in spectral optical data. Hotter (O/B/A) companions outshine the hot subdwarf at (almost) all wavelengths, whereas cooler companions (late M-type) contribute too little to be detectable even in the IR.

We used the SED fitting method described by Heber et al. (2018) to identify composite colour binaries and constrain the nature of the cool companions.[10] To model the two stellar components in the SED, we made use of two model grids. The hot components were modelled using our standard grid computed with Atlas12 (Kurucz 1996; Irrgang et al. 2018), which covers the full hot subdwarf parameter space and has been used extensively for many analyses of sdB stars, for example by Schaffenroth et al. (2022). For the MS component, a grid of synthetic PHOENIX stellar atmosphere models was used (Husser et al. 2013) which can be downloaded from the Göttingen Spectral Library.[11] The best fit to the observed SED was achieved using $\chi^2$ minimisation. Free fit parameters were the angular diameter ($\theta$) of the hot subdwarf, the effective temperature of both components, the surface ratio $R^2_{cool}/R^2_{hot}$, as well as the interstellar reddening parameterised by the colour excess $E(44-55)$. Here, we used the reddening law of Fitzpatrick et al. (2019). Because the surface gravity and metallicity of both components cannot be determined from the SED, we had to fix these values. The companion's surface gravity was fixed to 4.3, a reasonable value for possibly evolved MS companions. Their metallicity was fixed to solar. In the case of the hot subdwarf, the metal abundances were fixed to average sdB values from Pereira (2011) and the surface gravity was constrained to spectroscopic values where possible.

Of the spectroscopically identified hot subdwarfs in the sample, including those manually added systems, 48 (16%) are composite-colour binary systems (see Fig. 13 for several example fits). The majority are cooler G-, and K-type companions. Furthermore, we use the derived temperature of the companion and interpolate between table 15.7 given in Cox (2000), to discern its spectral class. A total of 16 systems have been newly identified, including 5 that were previously categorised as single hot subdwarfs. This fraction is somewhat lower than has previously been found. Work done by Stark & Wade (2003) identified composite-colour hot subdwarf binaries by means of colour-colour diagrams including IR fluxes from 2MASS and arrived at an estimate of 30% after approximating a volume-limited sample. The works of Thejll et al. (1995), Ulla & Thejll (1998), and Girven et al. (2012) all contributed to the current list of composite-colour hot subdwarf star binaries identified using colour-colour diagrams. Schaffenroth (2016) systematically constructed SEDs of hot subdwarf stars of various different natures from 5 different samples, arrived at the same estimate of 30%. These fractions are corroborated by the most recent study by Solano et al. (2022) who estimated 25%. Our lower estimate is likely due to the MS rejection criterion. Although several systems located within the MS were added manually, there remains a strong observational bias against detecting these systems as they are currently mainly discovered through UV excess rather than infrared excess or spectroscopic observation (Kawka et al. 2015) owing to the similar spectral coverage and large luminosity difference of the companion to that of the hot subdwarf. The impact of these missing systems will be discussed in Sect. 6 when assessing the completeness of the sample.

Lastly, the SEDs for those sources that failed our RUWE < 7 quality cut were inspected to search for potentially genuine composite hot subdwarf binaries, as well as potentially removed single sources. Barring GALEXJ110733.7-454424, a newly discovered composite sdB+G/K that was excluded from the sample due to a poor astrometric solution (see Sect. 2.2), all sources with RUWE > 7 are consistent with single MS stars. Among these, 21 have spectra in the LAMOST DR8 archive and exhibit the spectral features characteristic of MS A-type stars.

By combining the SED method with the *Gaia* parallax measurements and a spectroscopic analysis it would also be possible to derive radii, masses, and luminosities. Such an analysis will be performed in a follow-up paper once the spectroscopic analysis is complete. The same routine as described above is used in Sect. 6.2 to generate mock SED fits in our attempt to estimate the number of missing composite-colour hot subdwarfs in our sample.

*5.4. Cataclysmic variables*

Barring the hot subdwarf population, the CVs are the next most numerous objects in our selection with a count of 41 (10%), which includes three newly discovered systems: Gaia DR3 5389717630410364160, PB5919, and TYC7791-1293-1. Their presence in the sample stems from the fact that they photometrically fall in close proximity to the hot subdwarfs in the HRD, with a maximum density in the region $G_{BP} - G_{RP} \approx 0.56$ and $G_{abs} \approx 10.5$ (Abril et al. 2020). Moreover, the subtypes of intermediate polars (IPs) and nova-likes concentrate close to $G_{BP} - G_{RP} \approx 0.37$ and $G_{abs} \approx 5.63$ due to these stars being optically dominated by their bright accretion disks. Recently, Pala et al. (2020) compiled a volume-limited sample of CVs which constituted a total of 42 members within 150 pc. Since our sample is optimised for the selection of hot subdwarf stars, only two of the CVs listed in (Pala et al. 2020) are in our sample: IX Vel and CD-42 14462, both indeed Nova like systems which demonstrate strong hydrogen emission lines, indicative of a bright accretion disk. Given the parameter space our selection encapsulates in the HRD, the presence of these systems in our sample is consistent with what is expected (Abril et al. 2020). The remaining 39 systems are primarily too faint to meet our absolute magnitude selection, or are otherwise located within the MS.

*5.5. WDs, post-AGBs, and CSPNe*

A small fraction of stars in our sample are hot and bright WDs at the beginning of their cooling track. In the bottom left corner of Fig. 11 a clustering of WDs can be seen below the hot subdwarf cloud. As they are bright, all are well known barring two newly discovered and analysed systems in Reindl et al. (2023) - J17514575+3820157 and J194511.31−445954.57. GD 1532, currently listed as a known sdB in the SIMBAD Astronomical Database (Wenger et al. 2000), has been reclassified as a WD in this work. The population of WDs in our sample also includes some identified CSPNe which can be found in both the Hong Kong/AAO/Strasbourg H$\alpha$ (HASH) PN catalogue (Parker et al. 2016) and that of Chornay & Walton (2021), such as NGC 7293 and M27 which host DAO WDs in their cores (Giannias et al. 2011). Several of these post-AGBs and CSPNe, however, have bright and luminous sdO cores, such as PNA6636, NGC 1360, NGC 1514, and MWP 1 (Greenstein & Minkowski 1964; Mendez & Niemelä 1977; Feibelman 1997). To be in line with the nature of this works classification scheme as outlined in Sect. 3.3, those identified CSPNe or post-AGBs with spectroscopically classified sdO cores, are counted among the sdO stars for the analysis and statistics of our sample. Those CSPNe, as well as the PG 1159 stars with identified WD cores, will be counted among the spectroscopically identified WDs. KPD0005+5106, however, for example, is a known pre-WD object (Werner & Rauch 2015) with O(He) spectral type whose origin is still de-bated. It will be counted among the helium sdOs in this work. On this note, it is important to emphasise that many bright sdO stars share the same properties of several CSPNe (e.g. Giannin-nas et al. 2010) despite having very different evolutionary histories. Including this population in future monitoring programmes in the context of volume-complete samples of hot subluminous stars will certainly be of interest to help better disentangle the outcomes of both single and binary stellar evolution theory.

---

[10] This SED fitting routine was initially developed by Andreas Irrgang.
[11] http://phoenix.astro.physik.uni-goettingen.de/

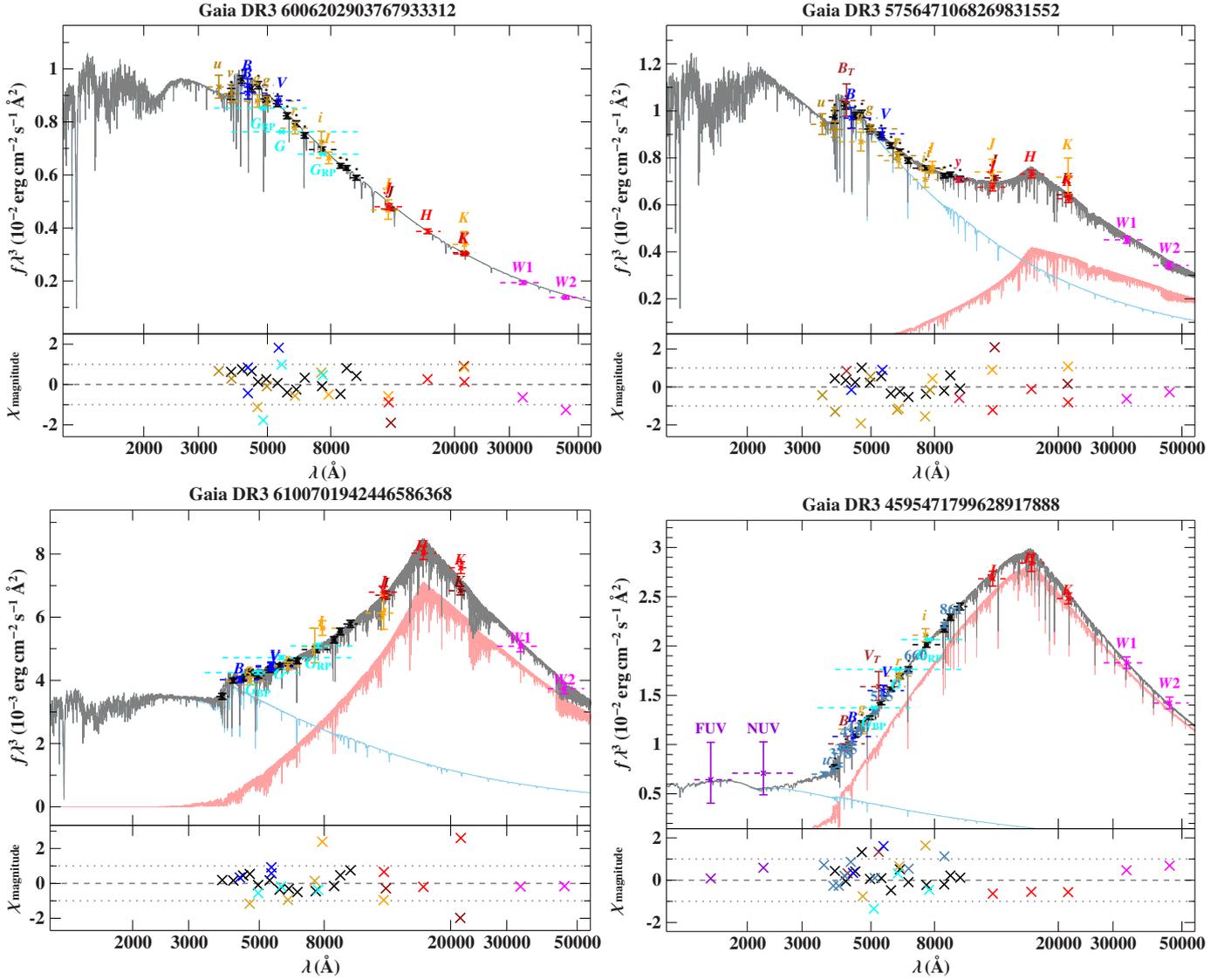

Fig. 13: Example spectral energy distribution fits for a single sdOB (top left), an sdB+M0 (top right), sdB+G9V (lower left), and a sdB+F6V (lower right). Photometric fluxes are shown as coloured data points with their respective filter widths as dashed horizontal lines. The grey line shows the combined model spectrum, while the individual contributions of the hot subdwarf and MS companion are given by the blue and red lines, respectively. The different photometric systems in these plots are assigned the following colours: JPLUS (steel blue; von Marttens et al. 2024), Gaia XP (black; Gaia Collaboration 2022), Gaia G/BP/RP (cyan; Gaia Collaboration 2022), WISE (magenta; Cutri et al. 2021), SkyMapper (dark gold; Onken et al. 2019), SDSS (gold; Alam et al. 2015), GALEX (dark violet; Bianchi et al. 2017), Pan-STARRS (crimson; Chambers et al. 2016), VISTA (maroon; Cioni et al. 2011; McMahon et al. 2013), DENIS (orange; DENIS Consortium 2005), 2MASS (red; Cutri et al. 2003), Johnson (blue; Henden et al. 2016; Girard et al. 2011) and Tycho (brown; Høg et al. 2000). Citations for all photometric data used in this work can be found in the acknowledgements.

### 5.6. Other objects of diverse and ambiguous classifications

There are 30 (10%) stars identified in this sample as late B- or early A-type stars, indicating that they are considerably cooler than sdB/O stars. They are probably either main sequence, blue horizontal branch stars, or show sdA-type spectra, which are characteristic of many different evolutionary classes (Pelisoli et al. 2017). Although most BHBs would be situated below the main sequence in the CMD, confirming this classification is difficult with the majority of the obtained spectra being low resolution. However, we couple a preliminary analysis of these spectra to constrain the effective temperature and surface gravity with our SED fitting routine (Sect. 5.3) to obtain the stellar radii and masses of these stars. We find ten to be consistent with MS stars; they are indicated accordingly in Fig. 11. Another eight have radii between 0.5 and 1.0 $R_\odot$, respectively, making MS an unlikely evolutionary class for these stars. Identifying them as probable BHB candidates, all of these objects are situated in close proximity to the blue horizontal branch in the CMD of Fig. 11 between the hot subdwarf cloud and the MS. Five stars have RUWE values greater than 4, which we deem to be an unreliable astrometric solution and are classed as MS/BHB. One source in our sample is a known Herbig Ae star: UY Ori A (Fairlamb et al. 2015).

[PS72]97, Gaia DR3 2513538251735261696, Gaia DR3 4318061098980872960, and Gaia DR3 6058834949182961536 are four objects in our sample that demonstrate MS A-type-like spectra, but have absorption lines that are too broad to be classified as a MS star. It is probable that these systems are helium-core white dwarf progenitors passing through our parameter space towards the WD cooling track and indeed [PS72]97 and Gaia DR3 2513538251735261696 have recently been identified as such (Kosakowski et al. 2023). Their positions in the CMD indicated as gold triangles in Fig. 11 are clearly grouped into a specific parameter range which is well separated from the hot subdwarf stars. These objects are accordingly classed as sdAs in this paper and will be subject to a more detailed spectroscopic analysis to disentangle their evolutionary origin in the future.

Lastly, two systems, TYC3135-86-1 and HD337604, have spectra resembling B-type MS stars. These stars have effective temperatures close to 20,000 K, but radii of 0.35±0.02 and 0.42± 0.02 $R_\odot$, respectively. This places them above the cool end of the EHB and below the MS. We class these two cases as pre-ELM/sdB candidates and indicate them as red stars in Fig. 11. These stars require follow-up observations to discern their true natures.

## 6. Completeness of the 500 pc sample

A completeness estimate is an essential property for any sample of stars that aims to test physical theories about stellar evolution. While we have demonstrated statistical homogeneity when accounting for the scale height, this alone cannot serve as a completeness estimate. The non-negligible influence of the Galactic disk structure at the 1 kpc scale, which corresponds to the diameter of our sample, necessitates a qualitative exploration of completeness in this section.

### 6.1. Reddening by interstellar extinction

The completeness of CG22, the source of this sample, is estimated to be 80 − 90% for single sources and unresolved binaries overall, with near completeness up to 1.8 kpc above and below the Galactic plane. However, within the Galactic plane, interstellar extinction-induced reddening can shift a source to redder colours, potentially moving it out of our selected region in the CMD. To account for this, we obtained reddening estimates of the form $E(44-55)$ for all non-composite hot subdwarfs using a 2D line-of-sight SED fitting routine, as detailed in Sect. 5.3. Where available, literature parameters of $T_{eff}$, $\log(g)$ and $\log(y)$ were used to constrain the fit, where $T_{eff}$ was otherwise left as a free parameter and $\log(g)$ and $\log(y)$ were fixed to typical atmospheric hot subdwarf parameters. For the composite-colour hot subdwarfs, however, we use the reddening given by the 3D Stilism[12] (Lallement et al. 2014; Capitanio et al. 2017) reddening maps because atmospheric parameters for these sources were often unavailable. Without precise atmospheric parameters of all targets to constrain the SED fitting procedure, the E(44-55) parameter can become unreliable due to a degeneracy between temperature and extinction.

The lower panel of Fig. 14 presents a CMD illustrating the shift each hot subdwarf undergoes when reddening is considered. Blue stars represent non-composites corrected for $E(44−55)$, while magenta stars depict composites corrected using Stilism. The impact of this correction is indicated by black-dotted lines, illustrating the shift in each source in this parameter space. The upper panel provides a colour-scaled skyplot, revealing the directional dependency of reddening, with the largest values concentrated towards the Galactic centre, as expected.

In the CMD parameter space, a typical non-composite hot subdwarf would require a reddening of ≈ 0.6 mag to fall outside our selection. The average reddening towards the Galactic centre (covering about 17% of the sky with $|b| < 30°$ and $300° < l < 60°$) for our targets in the 500 pc sample, according to Stilism reddening maps, is ≈ 0.11 mag. We argue that reddening has a negligible impact on the completeness of the single sources and unresolved binaries, and the sample should be essentially complete up to 500 pc across the sky. An exception is UCAC4 436-075435, an sdOB with a calculated reddening of $E(44−55)$ = 0.605 mag. While this system initially appears in the bottom right corner of the CMD in Fig. 14, it aligns with the centre of the hot subdwarf cloud once reddening is corrected. As no other non-composite hot subdwarf in our sample exceeds a reddening of 0.3 mag, we anticipate that reddening does not significantly affect the completeness of our sample.

It is important to note that the population of low-mass hot subdwarf stars or helium-core WD progenitors such as HD 188112 typically occupy a region closer to the MS and are probably highly incomplete in our sample. Given their low occurrence rate in our selected parameter space, we do not account for them here and assume completeness for all single sources and hot subdwarf binaries which host companions that have little impact on the overall colour of the system, such as WD or cool MS or BD companions. However, incompleteness due to extinction will become an increasing issue for future volume-limited samples that extend beyond 500 pc.

The situation is different for the composite binary hot subdwarfs, where the MS companion inherently shifts the system to redder colours in the *Gaia* CMD parameter space, irrespective of interstellar extinction. Assuming clean astrometry, the companion's presence increases the overall brightness of the system, shifting its position in the CMD upwards instead of downwards. This creates a trend of increasing absolute magnitude with increasing colour index. The extent of this shift is not straightforward. In the simplest case, the shift is proportional to the spectral type of the MS companion, with earlier types being brighter than later types, and their brightness exponentially more impactful than their colour. However, the exact evolutionary status of both the hot subdwarf and MS is unknown, and the relative contribution of each to the system's flux is uncertain, influencing the system's CMD position.

### 6.2. The missing composite-colour hot subdwarfs

To estimate the completeness of our sample and identify potential missing composite systems, we employed a theoreti-cal approach. Mock binary SEDs were generated, and resulting colours and absolute magnitudes were computed for various combinations of component parameters. MS companion masses ranged from 0.25 $M_\odot$ to 3.0 $M_\odot$ at 0.05 $M_\odot$ intervals, considering three types of hot subdwarfs: sdB, sdOB, and He-sdO. Typical atmospheric parameters were assumed for each type of hot subdwarf based on our sample. Companion parameters were calculated from their intermediate age, defined as the point where central hydrogen abundance reaches 30% with a threshold for effective temperatures below 11,000 K (see Heber et al. 2018, for the fitting routine details). Figure 15 displays the results of these calculations, showing good agreement with the identified composites in our sample. An exception is NGC 1514 in the upper right corner, hosting a spectral type A0 horizontal-branch star companion at the centre of a planetary nebula. BD−03 5357 and HD 185510, hosting subgiant companions, are not included in this plot due to plot limits. Companion masses of 1.1, 1.0 and 1.0 for the He-sdO's, sdB's, and sdOB's, respectively, lie closest to the MS rejection criterion. Accounting for the effects of reddening due to interstellar extinction with the average Stilism value near the Galactic centre for our sources ($\approx 0.11$ mag), the detectable companion masses are decreased by $\approx 0.1$ $M_\odot$ each.

---

[12] https://stilism.obspm.fr/

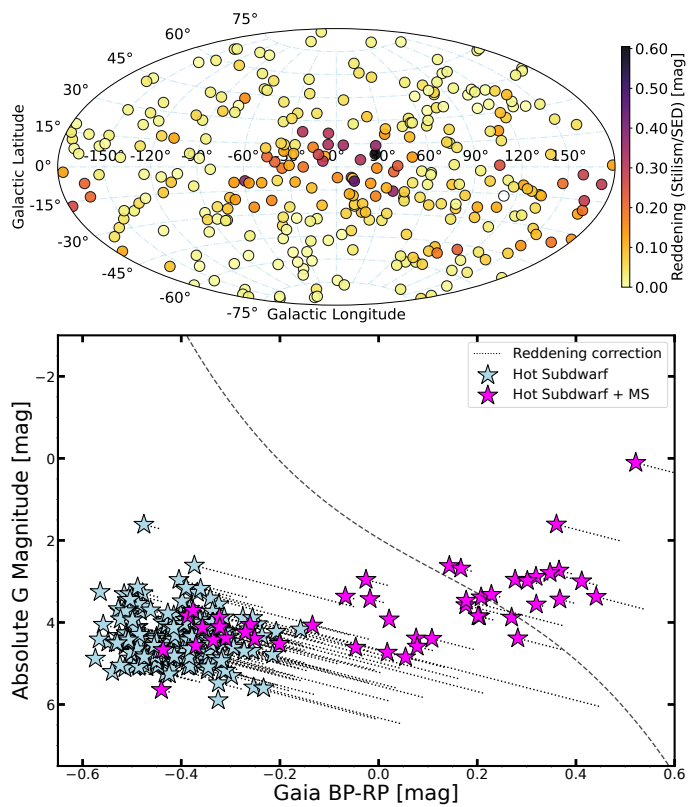

Fig. 14: Plot to illustrate the effect of interstellar extinction on the 500 pc sample. Top: Skyplot of the 500 pc sample which is colour-coded according to the reddening parameters given by both Stilism and the SED fits. Bottom: $G_{BP} - G_{RP}$ vs $M_G$ diagram of the same sources where blue stars indicate non-composite hot subdwarfs that have been corrected for interstellar extinction using the E(44-55) SED estimates, and magenta stars indicating composite-colour hot subdwarfs that have been corrected for using Stilism. The correction shift is shown with the black dotted lines. Systems hosting subgiant companions have been excluded from the plot.

Corrected for the relative abundances of each hot subdwarf class in our sample, this yields an average companion mass of 0.91 $M_\odot$ - the theoretical maximum observable mass in our selection. Consequently, our sample is expected to be, on average, complete for companions of spectral type G5 or later across the entire sky. This estimate is supported by the fact that all manually added systems in our sample host companions of spectral types G or earlier. However, it remains challenging to estimate the number of missing composites without assuming an underlying mass distribution for the MS companions.

Stark & Wade (2003), approximating a volume-limited sample, derived a composite-colour hot subdwarf fraction of 30%, while a more recent study by Solano et al. (2022) arrived at 25%. Taking the average of these two studies, our initially observed fraction of 16% (48 systems) can then be adjusted to 27.5%, suggesting that $\approx 31$ systems may be missing from our selection. This estimate probably represents a lower limit, considering biases against composite-colour binaries in previous samples of known hot subdwarf stars. Consequently, we infer a 90% overall completeness for the hot subdwarf star constituents, assuming a homogeneous distribution of composite-colour hot subdwarfs. To account for these estimated missing systems, we add the relative error of the 31 systems to the upper uncertainty on the derived mid-plane density, resulting in $\rho_0 = 6.15^{+1.16}_{-0.53} \times 10^{-7}$ stars/pc$^3$.

We now take an observational standpoint to predict the number of missing composites in our selection by cross-matching a *Gaia* database query with GALEX. Hot subdwarfs in unresolved binary systems with an F or G-type MS counterpart should show an excess in the ultraviolet (UV) (see Downes 1986, for early works). The selection criterion for the executed query to the *Gaia* database, and the defined UV excess for sources with GALEX detections are given in table 4.

| 1. Selection criterion for the queried region shown in Fig. 15. |
|---|
| $G_{abs} < 17.7(G_{BP} - G_{RP})^3 - 6.9(G_{BP} - G_{RP})^2 + 7.35(G_{BP} - G_{RP}) + 1.95$ |
| $< 7 - 3(G_{BP} - G_{RP})$ |
| $> 7(G_{BP} - G_{RP}) - 3.8$ |
| $> -1$ |
| RUWE < 7 |
| parallax_over_error > 5 |
| 2. UV excess criterion for queried sources with GALEX data. |
| FUVmag − NUVmag > 2.5 − 1.5(NUVmag − Gmag) |

Table 4: Selection criteria.

The queried area, encompassed by the black dashed lines in Fig. 15, accounts for a reddening of up to 0.3 magnitudes as well as the full range of masses of the companions to hot subdwarfs which were calculated from our mock SEDs. The query does not cover the parameter space occupied by the hot subdwarfs hosting subgiant companions such as BD−03 5357 and HD185510. This query yielded 1,003,196 sources, with 516,902 having GALEX detections (51.5%). Among these, 31 exhibit UV excess as depicted in Fig. 16, and within this subset, 21 are unidentified candidates. Based on the number of selected sources in GALEX compared to *Gaia* - 51.5% - that would suggest that $\approx 42$ candidate systems are within the MS given our selection criteria. This number is in reasonable agreement with our estimate given above, especially if we account for a number of WDs hosting MS companions which would be indistinguishable using this method. Spectroscopic follow-up and RV monitoring of these candidates are needed in order to pursue this missing population.

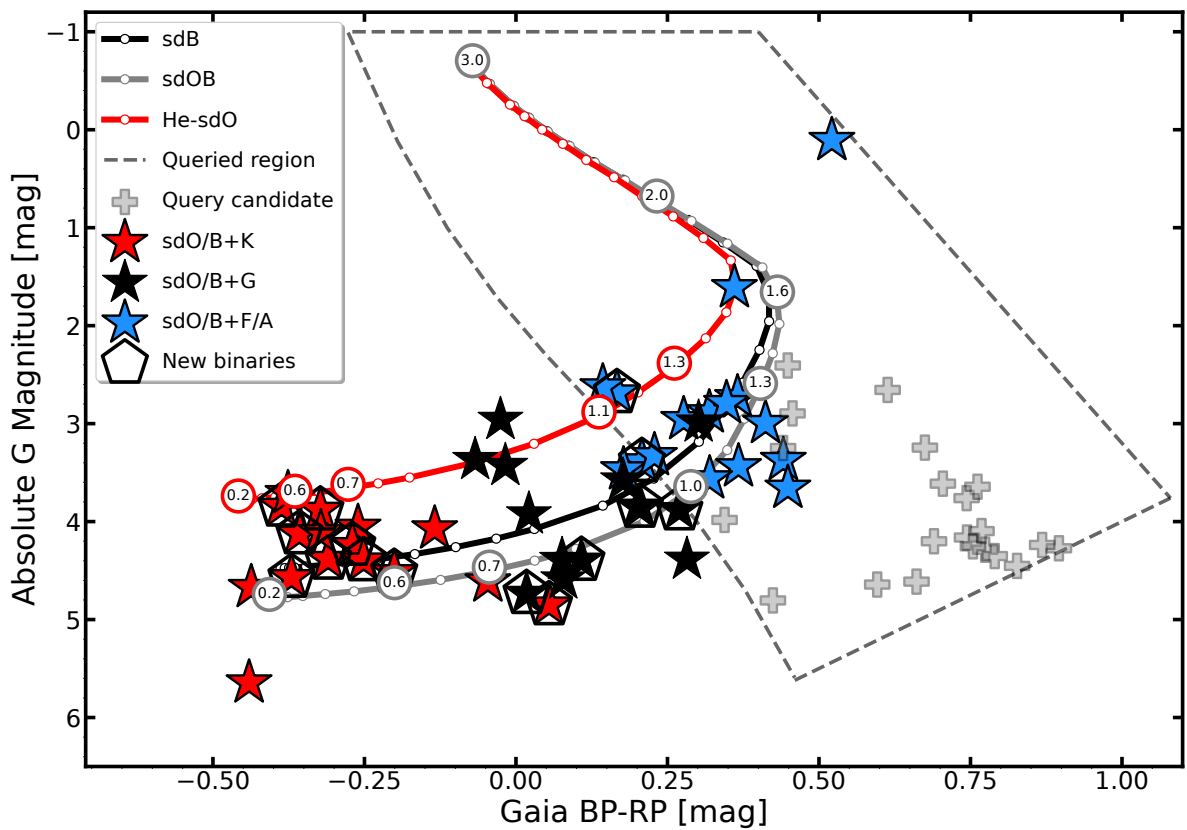

Fig. 15: CMD displaying the results of the mock SED fits for different combinations of hot subdwarf (sdB-black; sdOB-grey; He-sdO-red) with MS companions. Each circle along the tracks indicates a 0.05 $M_\odot$ step in the companion mass between 0.2 $M_\odot$ and 3.0 $M_\odot$ where several have been labelled as reference points. These tracks indicate that a companion mass of 1.0 - 1.1 $M_\odot$ is the maximum expected to be detected in this sample. The calculations do not account for reddening and can be compared to the reddening-corrected hot subdwarf composite systems given in the plot. These composites have been divided up into companion types to emphasise that the sample is expected to be complete for all K-type companions. Newly discovered systems are indicated with an additional encompassing pentagon. BD−03 5357 and HD185510 are outside the plot limits; they are known to host subgiant companions. Queried candidates that will be followed-up in a spectroscopic monitoring campaign are shown as grey plus signs.

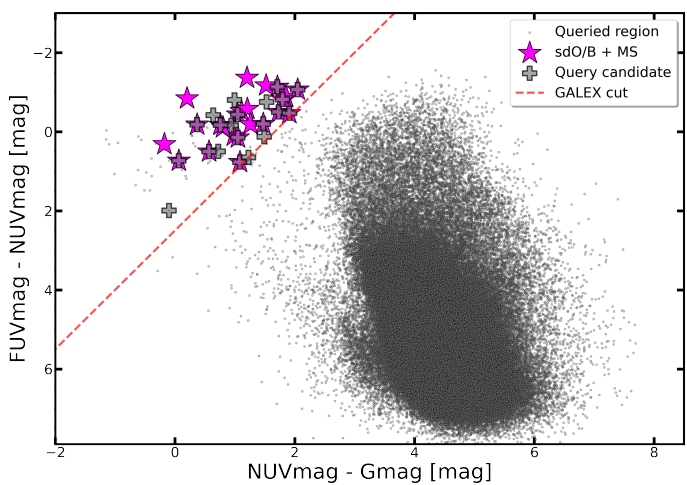

Fig. 16: Colour-colour diagram displaying the 516,902 queried sources with GALEX detections. The red dashed line delineates our UV-excess cut, and to its left are situated all known composite hot subdwarfs (magenta stars) within 500 pc. The 21 promising composite candidates are represented by grey plus signs.

### 6.3. Hot subdwarfs as companions to Be stars

Be stars are rapidly rotating B-type stars, either MS or evolved, which have shown some form of emission lines at some point during their lives (Porter & Rivinius 2003). Recent observational evidence has supported the formation of early-type Be stars to be rooted in binary interaction (Bodensteiner et al. 2020), where the rapid rotation and decretion disk is a result of a previous episode of mass transfer from a nearby companion. If this were true, then many Be stars should host the remnants of the donor star. Indeed, several evolved compact companions to Be stars are known, including a handful of stripped helium burning hot subdwarf stars which are predicted to make up the majority of the companions (Pols et al. 1991). Due to their much higher intrinsic luminosity ($\approx 100$ times) and similar spectral energy range, Be stars can effectively hide a companion hot subdwarf in the optical sector. Ultraviolet (UV) data is required to distinguish these components which has been the method of detection of most previous studies. More recently, El-Badry et al. (2022) performed a dedicated search for bloated stripped companions to Be stars using data from the APOGEE survey. From a well-defined parent sam-ple of 297 Be stars, one mass-transfer binary was identified, and predicted to eventually evolve into a Be + sdO binary system. Because the timescale of this mass transfer phase is predicted to be fleeting ($\approx$ 1 Myr), the authors estimate that $(10-60)\%$ of Be stars host stripped companions. The BeSS database offers an up to date count of all the known Be stars, as well as offering high-quality spectra for most sources. By querying the archive, we find 305 Be stars within 500 pc. Following the prescription of El-Badry et al. (2022), this would imply that there are $28-170$ hot subdwarf stars hiding behind classical Be stars in our 500 pc sample.

Analysing a sample of 13 Be stars using Hubble Space Telescope/STIS far-ultraviolet spectra, Wang et al. (2021) identified the presence of a hot sdO companion in ten of these stars. Notably, seven of these companions are located within 500 pc. These systems are exceedingly difficult to detect, especially since even the brightest sdO companions may only contribute $\approx 1\%$ of the total flux ratio (Peters et al. 2016). Hot subdwarfs around Be stars have probably been formed by stripping through binary interactions, but tend to possess larger masses and radii than the general population discussed in this work. This stems from an entirely different evolutionary channel for binary stars of high mass. To this end, we refrain from including them in the analysis presented in this work, and leave it as an open question for future studies as to how they may be integrated into the population of known hot subdwarf stars.

## 7. The birthrate of hot subdwarf stars

In Sect. 4, we determined the mid-plane density of the non-composite hot subdwarf stars in our sample to be $\rho_0 = 5.17 \pm 0.33 \times 10^{-7}$ stars/pc$^3$, which increases to $\rho_0 = 6.15^{+1.16}_{-0.53} \times 10^{-7}$ stars/pc$^3$ when considering the known composite-colour hot subdwarfs within 500 pc and accounting for incompletness (see Sect. 6).

To estimate the birthrate, we rely on theoretical evolutionary lifetimes from Han et al. (2003). Interpolating between their $0.5 M_\odot$ and $0.45 M_\odot$ models, we derive a value of $\tau = 1.98 \times 10^8$ years, adjusted with a 10% increase to include the helium shell burning phase. Employing a Monte Carlo approach, assuming an error of 50 Myr on the hot subdwarf lifetime, our calculated median birthrate is $\rho_0/\tau = 3.35^{+1.24}_{-0.77} \times 10^{-15}$ stars/pc$^3$/yr$^{-1}$ for our sample. Assuming a Galactic volume of $5 \times 10^{11}$ pc$^3$ (Zombeck 1990), Han et al. (2003) give a prediction of $10 \times 10^{-14}$ stars/pc$^3$/yr$^{-1}$ for their best-fit model, which is $\sim 30$ times higher than our observational estimate, even after accounting for our attempt to correct for the incompleteness of composite binaries. The theoretical estimate halves when this same selection effect is accounted for in the models, although is still on the same order of discrepancy as stated above when composites are excluded from our sample. The precise fraction of composites among all hot subdwarfs would be a very decisive number to distinguish simulation sets (see Table 2 of Han et al. 2003). Unfortunately it is difficult to measure this fraction as discussed throughout this paper.

While these estimates may appear conflicting, it is essential to note that they might represent slightly different quantities. This work provides a comprehensive count of all hot subdwarfs within this parameter space, without distinguishing, for instance, the pure horizontal branch (HB) evolution considered by Han et al. (2002, 2003). Our sample, representing the general local population of hot subdwarf stars in the solar neighbourhood, differs from the approach taken by Han et al. (2003), whose birthrate calculations apply to the entire Galactic population. Notably, our local volume-limited sample under-represents the thick-disk population mentioned by Han et al. (2003), where a higher birthrate is predicted compared to population I stars under similar parameters. Despite potential overestimations in our sample completeness, the significant discrepancy remains challenging to explain, particularly given our density estimate focusing on the mid-plane, where the highest concentration of stars is expected. Further characterisation of this sample may give clues about the underlying reason for this miss-match. For instance, refinement of the BPS models may be sought by observationally constraining the relative number of systems with MS and WD companions, which would help to constrain model input parameters and thus shed insight on the relative importance of each of the formation channels. This ambition, will be the main focus of a future paper.

## 8. Summary

The creation of this all-sky volume-limited sample, defined using the accurate parallax measurements from *Gaia's* DR3, is the first of its kind. It has contributed to the identification and classification of 83 new hot subdwarfs, adding to a total of 305 well-classified or otherwise updated hot subdwarf systems within 500 pc. Among these, 48 systems (16%) exhibit infrared excess in their SED fits, and they have been categorised as composite systems hosting MS companions. Notably, 16 of these composites are newly identified in this study, primarily comprising systems with K- or G-type companions, which are more challenging to detect through optical spectroscopy compared to companions of type F, dominating the optical data. Additionally, the sample revealed three nova-like CVs closely situated to the hot subdwarfs in the *Gaia* colour-magnitude parameter space. Throughout the observing programme, over 300 A-type MS or BHB stars were also observed, the majority of which were previously unknown. For the hot subdwarf stars, we estimate an overall completeness of 90%, with composite hot subdwarfs assumed to be the main contributors to the deficit. By segmenting the sample along the Galactic $z$-axis, we model the non-composite hot subdwarf population, finding the best fit with a hyperbolic secant function. The simultaneous fit parameters, mid-plane density ($\rho_0$) and the scale height (h$_z$), are determined as $5.17 \pm 0.33 \times 10^{-7}$ stars/pc$^3$ and $281 \pm 62$ pc, respectively. When accounting for our completeness estimate and the inclusion of composites, assuming a homogeneous distribution, $\rho_0$ increases to $\rho_0 = 6.15^{+1.16}_{-0.53} \times 10^{-7}$ stars/pc$^3$. While this number is lower than theoretical estimates, it is representative of the local disk population in the solar neighbourhood. Limitations of the fitting procedure include low-number statistics and non-homogeneous Galactic structure across our sample, leading to increased uncertainties, particularly in the scale height. Despite these challenges, our sample is statistically consistent with a homogeneously distributed sample in 3D space when accounting for the scale height, suggesting it as the dominant factor contributing to the varying density across the sample. Our volume-limited sample reveals novel number distributions among hot subdwarf subclasses, including a higher fraction of sdOBs compared to sdBs, a distinction not previously unambiguously identified. However, the true fraction of composite-colour hot subdwarfs remains challenging and is an ambition for future studies. Additionally, we find the commonly adopted selection criterion of AEN < 1 and RUWE < 1.4 is sub-optimal for our sample, especially for the composite hot subdwarf binary population. As an alternative, we propose empirical thresholds to assess the as-trometric solution quality for stars within 500 pc from the *Gaia* database.

Interest in spectroscopically complete, volume-limited samples can only increase in the future. With *Gaia* continuing to offer integral advancements in its unprecedented astrometric solution with every installment, as well as upcoming large-scale spectroscopic surveys like WEAVE (Jin et al. 2023) and 4MOST (de Jong et al. 2019), we can continue to expand our samples at an ever increasing rate. The need for an up-to-date public database, which we continue to work on,[13] is essential, especially if we wish to better organise our knowledge as a community, promote new studies, and drive progress towards ever greater insights onto the nature of hot subluminous stars in our Galaxy.

*Acknowledgements.* We would like to thank all those involved in the spectroscopic follow-up campaign that has ensued over the recent years. H. D. is and N. R. was supported by the Deutsche Forschungsgemeinschaft (DFG) through grant GE2506/17-1. NR is supported by the DFG through grant RE 3915/2-1. V. S. and M. P. received funding by the Deutsche Forschungsgemeinschaft (DFG) through grants GE2506/9-1 and GE2506/12-1. D.S. acknowledges funding by DFG grant HE1356/70-1. I. P. acknowledges funding from a Warwick Astrophysics prize post-doctoral fellowship, made possible thanks to a generous philanthropic donation, and from a Royal Society University Research Fellowship (URF\R1\231496). JV acknowledges support from the Grant Agency of the Czech Republic (GAČR 22-34467S). The Astronomical Institute Ondřejov is supported by the project RVO:67985815. M. U. gratefully acknowledges funding from the Research Foundation Flanders (FWO) by means of a junior postdoctoral fellowship (grant agreement No. 1247624N). T. S. acknowledges funding from grant SONATA BIS no 2018/30/E/ST9/00398 from the Polish National Science Centre (PI T. Kamiński). RR acknowledges support from Grant RYC2021-030837-I funded by MCIN/AEI/ 10.13039/501100011033 and by "European Union NextGeneration EU/PRTR". This work was partially supported by the AGAUR/Generalitat de Catalunya grant SGR-386/2021 and by the Spanish MINECO grant PID2020-117252GB-I00. This work has made use of the BeSS database, operated at LESIA, Observatoire de Meudon, France: http://basebe.obspm.fr. Some data in this worked came from Guaranteed Observation Time (GTO) based on observations collected at the Centro Astronomico Hispano en Andalucia (CAHA) at Calar Alto, operated jointly by Junta de Andalucia and Cosejo Superior De Investigaciones Cientificias (IAA-CSIC). The research has made use of TOPCAT, an interactive graphical viewer and editor for tabular data Taylor (Taylor 2005). This research made use of the SIMBAD database, operated at CDS, Strasbourg, France; the VizieR catalogue access tool, CDS, Strasbourg, France. This work has made use of data from the European SpacebAgency (ESA) mission *Gaia* (https://www.cosmos.esa.int/gaia), processed by the *Gaia* Data Processing and Analysis Consortium (DPAC), https://www.cosmos.esa.int/web/gaia/dpac/consortium). Funding for the DPAC has been provided by national institutions, in particular the institutions participating in the *Gaia* Multilateral Agreement. T.K. acknowledges support from the National Science Foundation through grant AST #2107982, from NASA through grant 80NSSC22K0338 and from STScI through grant HST-GO-16659.002-A. Co-funded by the European Union (ERC, CompactBINARIES, 101078773). Views and opinions expressed are however those of the author(s) only and do not necessarily reflect those of the European Union or the European Research Council. Neither the European Union nor the granting authority can be held responsible for them. We thank the survey teams responsible for the copious amount of photometric data available in the literature which we utilised in this work (Henden et al. 2016; DENIS Consortium 2005; Onken et al. 2019; McMahon et al. 2013; Schlafly et al. 2019; Riello et al. 2021; Kato et al. 2007; Høg et al. 2000; van Leeuwen 2007; Girard et al. 2011; Mermilliod 2006; Rufener 1999; Hauck & Mermilliod 1998; Ducati 2002; Spitzer Science 2009; Lucas et al. 2008; Lawrence et al. 2007a,b; Cutri et al. 2021; Bianchi et al. 2017; Yershov 2014; Chambers et al. 2016; Shanks et al. 2015; Page et al. 2012; Abbott et al. 2018; Thompson et al. 1978; Morel & Magnenat 1978; Kilkenny et al. 1988; Lamontagne et al. 2000; Zaritsky et al. 2004; Landolt 2007; Greiss et al. 2012; Norris et al. 1999; Chen et al. 2020; Werner et al. 2021; Bowyer et al. 1995; Kilkenny et al. 1997; O'Donoghue et al. 2013; Kilkenny et al. 2015; Stetson et al. 2019; Flewelling 2018; Nidever et al. 2021; von Marttens et al. 2024; Alam et al. 2015; Monguió et al. 2020; Drlica-Wagner et al. 2022; Abbott et al. 2021; Dixon & Kruk 2009; Gaia Collaboration 2022; Schlafly et al. 2018; Mendes de Oliveira et al. 2019; Edge et al. 2013; Cioni et al. 2011; Minniti et al. 2010; Wamsteker et al. 2000; Meixner et al. 2006).

---

[13] https://a15.astro.physik.uni-potsdam.de/w/the-hot-subdwarf/systems/stars/

[1] Institute for Physics and Astronomy, University of Potsdam, Karl-Liebknecht-Str. 24/25, 14476 Potsdam, Germany
e-mail: hdawson@astro.physik.uni-potsdam.de
[2] Dr. Remeis-Sternwarte and ECAP, Astronomical Institute, University of Erlangen-Nürnberg, Sternwartstr. 7, D-96049 Bamberg, Germany
[3] Department of Physics, University of Warwick, Gibet Hill Road, Coventry CV4 7AL, UK
[4] Thüringer Landessternwarte Tautenburg, Sternwarte 5, D-07778 Tautenburg, Germany
[5] Zentrum für Astronomie der Universität Heidelberg, Landessternwarte, Königstuhl 12, 69117, Heidelberg, Germany
[6] Instituto de Física y Astronomía, Universidad de Valparaíso, Gran Bretaña 1111, Playa Ancha, Valparaíso 2360102, Chile
[7] European Southern Observatory, Alonso de Cordova 3107, Santiago, Chile
[8] Astronomical Institute of the Czech Academy of Sciences, CZ-251 65, Ondřejov, Czech Republic
[9] Department of Astrophysics/IMAPP, Radboud University, P O Box 9010, NL-6500 GL Nijmegen, The Netherlands
[10] Max Planck Institut für Astrophysik, Karl-Schwarzschild-Straße 1, 85748 Garching bei München, Germany
[11] Recogito AS, Storgaten 72, N-8200 Fauske, Norway
[12] Nordic Optical Telescope, Rambla José Ana Fernández Pérez 7, ES-38711 Breña Baja, Spain
[13] Universitat Politècnica de Catalunya, Departament de Física, c/ Esteve Terrades 5, 08860 Castelldefels, Spain
[14] Steward Observatory, University of Arizona, 933 North Cherry Avenue, Tucson, AZ 85721, USA
[15] Leibniz-Institut für Astrophysik Potsdam (AIP), An der Sternwarte 16, 14482 Potsdam, Germany
[16] Armagh Observatory and Planetarium, College Hill, Armagh BT61 9DG, United Kingdom
[17] Instituto de Física, Universidade Federal do Rio Grande do Sul, 91501-900 Porto-Alegre, RS, Brazil
[18] Nordic Optical Telescope, Rambla José Ana Fernández Pérez, 7, ES-38711 Breña Baja, Spain
[19] Institute of Astronomy, KU Leuven, Celestijnenlaan 200D, B-3001 Leuven, Belgium
[20] Isaac Newton Group of Telescopes, Apartado de Correos 368, E-38700 Santa Cruz de La Palma, Spain
[21] Department of Physics and Astronomy, Aarhus University, Munkegade 120, DK-8000 Aarhus C, Denmark
[22] Department of Physics and Astronomy, University of Sheffield, Sheffield S3 7RH, UK
[23] Nicolaus Copernicus Astronomical Centre, ul. Rabiańska 8, 87-100 Toruń, Poland
[24] Anton Pannekoek Institute for Astronomy, University of Amsterdam, 1090 GE Amsterdam, The Netherlands
[25] Hamburger Sternwarte, University of Hamburg, Gojenbergsweg 112, D-21029 Hamburg, Germany

# Appendix A: Additional tables

Table A.1: 500 pc sample of non-composite hot subdwarf stars from *Gaia* DR3, sorted by increasing distance.

| Name | RAJ2000 | DECJ2000 | Class | G | BP-RP | Stilism E(B - V) | E(44-55) | Parallax | Parallax error | RUWE | Reference |
|---|---|---|---|---|---|---|---|---|---|---|---|
| HD188112 | 298.6311 | -28.3390 | sdB+WD | 10.17 | -0.32 | 0.00 | 0.00 | 14.002 | 0.059 | 1.04 | 2003A&A...411L.477H |
| HD149382 | 248.5972 | -4.0145 | sdOB | 8.89 | -0.44 | 0.00 | 0.03 | 13.272 | 0.064 | 1.20 | 1994ApJ...432..351S |
| BD+284211 | 327.7957 | 28.8637 | sdO | 10.45 | -0.54 | 0.02 | 0.00 | 8.917 | 0.081 | 1.18 | 2015A&A...579A..39L |
| BD+254655 | 329.9247 | 26.4324 | He-sdO | 9.65 | -0.47 | 0.03 | 0.01 | 8.679 | 0.062 | 1.22 | 1966VA......8...63G |
| HD171858 | 279.4860 | -23.1932 | sdB+WD | 9.80 | -0.35 | 0.13 | 0.05 | 7.245 | 0.050 | 0.79 | 2010A&A...519A..25G |
| BD+252534 | 189.3480 | 25.0665 | sdOB | 10.46 | -0.46 | 0.01 | 0.02 | 7.063 | 0.066 | 1.28 | 2006A&A...452..579O |
| BD+75325 | 122.7063 | 74.9661 | He-sdO | 9.49 | -0.53 | 0.01 | 0.00 | 6.879 | 0.063 | 1.30 | 1997ApJ...485..843L |
| HD191351 | 303.3812 | -65.0352 | sdB | 11.31 | -0.35 | 0.01 | 0.00 | 5.920 | 0.051 | 1.24 | 2013MNRAS.431..240O |
| PG0342+026 | 56.3940 | 2.7978 | sdB | 10.93 | -0.23 | 0.09 | 0.12 | 5.860 | 0.052 | 1.09 | 2013A&A...557A.122G |
| HD4539 | 11.8718 | 9.9822 | sdB | 10.26 | -0.38 | 0.01 | 0.02 | 5.828 | 0.079 | 1.47 | 2012MNRAS.427.2180N |
| BD-073477 | 191.0844 | -8.6714 | sdB+dM | 10.59 | -0.36 | 0.02 | 0.05 | 5.800 | 0.066 | 1.14 | 2015A&A...576A..44K |
| HD127493 | 218.0894 | -22.6572 | He-sdO | 9.96 | -0.43 | 0.04 | 0.05 | 5.721 | 0.064 | 1.03 | 2009PhDT.......273H |
| CD-321567 | 60.2723 | -32.3961 | sdB | 11.15 | -0.40 | 0.00 | 0.02 | 5.465 | 0.039 | 1.14 | 2013A&A...557A.122G |
| BD+393226 | 266.6331 | 39.3192 | He-sdO | 10.13 | -0.48 | 0.02 | 0.04 | 5.295 | 0.047 | 1.12 | 1981PhDT........113G |
| [CW83]0512-08 | 78.6829 | -8.8019 | sdOB | 11.18 | -0.45 | 0.01 | 0.04 | 5.278 | 0.063 | 1.36 | 2013A&A...557A.122G |
| TYC4406-285-1 | 215.3652 | 71.4058 | sdB | 11.22 | -0.39 | 0.01 | 0.00 | 5.239 | 0.038 | 0.90 | 2015MNRAS.450.3514K |
| [CW83]1758+36 | 270.0784 | 36.4823 | sdOB | 11.30 | -0.43 | 0.02 | 0.04 | 5.201 | 0.045 | 1.07 | 2013A&A...557A.122G |
| EC21494-7018 | 328.4220 | -70.0756 | sdB | 11.58 | -0.37 | 0.01 | 0.01 | 4.886 | 0.040 | 1.17 | 2012MNRAS.427.2180N |
| HD296955 | 144.5039 | -45.5410 | sdB | 10.92 | -0.35 | 0.03 | 0.00 | 4.879 | 0.037 | 1.04 | TW |
| TYC4101-97-1 | 100.5581 | 63.6663 | sdOB | 11.43 | -0.41 | 0.05 | 0.04 | 4.831 | 0.062 | 1.70 | TW |
| GALEXJ080510.9-105834 | 121.2951 | -10.9766 | sdB+WD | 12.23 | -0.28 | 0.01 | 0.05 | 4.780 | 0.053 | 0.96 | 2012MNRAS.427.2180N |
| CPD-311701 | 114.1257 | -32.2120 | He-sdO | 10.47 | -0.49 | 0.01 | 0.02 | 4.741 | 0.052 | 1.27 | 2009PhDT.......273H |
| PG1032+406 | 158.8186 | 40.3538 | sdB+WD | 11.45 | -0.46 | 0.01 | 0.01 | 4.730 | 0.062 | 1.29 | 2021ApJS..256...28L |
| CPD-691434 | 163.5406 | -70.2318 | sdB | 11.48 | -0.26 | 0.10 | 0.14 | 4.704 | 0.037 | 1.06 | TW |
| TYC4427-1021-1 | 256.2430 | 73.0785 | sdB | 10.73 | -0.41 | 0.02 | 0.00 | 4.683 | 0.040 | 1.23 | TW |
| TYC9044-1653-1 | 240.0493 | -64.5588 | He-sdB | 11.96 | -0.40 | 0.06 | 0.05 | 4.508 | 0.051 | 1.07 | 2012MNRAS.427.2180N |
| CPD-64481 | 86.9971 | -64.3843 | sdB+dM | 11.27 | -0.42 | 0.01 | 0.01 | 4.431 | 0.041 | 1.18 | 2015A&A...576A..44K |
| BD+48433 | 21.7668 | 49.7096 | sdB | 11.06 | -0.32 | 0.08 | 0.08 | 4.431 | 0.061 | 1.21 | TW |
| CD-2315853 | 298.9093 | -23.2285 | sdO | 11.68 | -0.36 | 0.08 | 0.12 | 4.416 | 0.057 | 1.21 | 1983ApJ...270L..13D |
| Feige34 | 159.9032 | 43.1024 | sdO+dM | 11.11 | -0.49 | 0.01 | 0.02 | 4.410 | 0.116 | 2.24 | 2018A&A...609A..89L |
| CD-24731 | 25.9527 | -24.0864 | sdB+WD | 11.68 | -0.49 | 0.01 | 0.01 | 4.392 | 0.069 | 1.38 | 2006A&A...452..579O |
| BD+481777 | 142.6955 | 48.2729 | He-sdOB | 10.64 | -0.49 | 0.01 | 0.02 | 4.256 | 0.069 | 1.21 | 2021ApJS..256...28L |
| HD319179 | 273.8647 | -31.7587 | sdB | 11.30 | -0.20 | 0.10 | 0.16 | 4.244 | 0.047 | 0.55 | TW |
| GALEXJ104148.9-073031 | 160.4534 | -7.5087 | sdB | 11.63 | -0.38 | 0.02 | 0.03 | 4.196 | 0.073 | 1.44 | 2012MNRAS.427.2180N |
| TYC6017-419-1 | 132.9954 | -17.2113 | sdOB | 11.78 | -0.41 | 0.05 | 0.06 | 4.183 | 0.062 | 1.28 | 2012MNRAS.427.2180N |
| CD-3515910 | 356.0916 | -34.4501 | sdB | 10.91 | -0.43 | 0.01 | 0.01 | 4.169 | 0.072 | 1.61 | 2012MNRAS.427.2180N |
| BD+423250 | 286.9189 | 42.3060 | sdOB | 10.59 | -0.33 | 0.02 | 0.06 | 4.151 | 0.036 | 1.20 | 2010A&A...513A....6O |
| HD205805 | 324.7947 | -46.0977 | sdB | 10.14 | -0.38 | 0.01 | 0.01 | 4.136 | 0.059 | 1.14 | 2006BaltA..15..107P |
| CD-55317 | 20.0538 | -55.2978 | sdOB | 11.94 | -0.47 | 0.01 | 0.00 | 4.060 | 0.046 | 1.01 | 2016MNRAS.459.4343K |
| TYC4000-216-1 | 353.7175 | 53.7841 | sdOB | 11.84 | -0.40 | 0.04 | 0.03 | 4.035 | 0.046 | 1.07 | 2015MNRAS.450.3514K |
| TYC4544-2658-1 | 145.9731 | 78.5281 | sdB | 11.11 | -0.42 | 0.01 | 0.01 | 3.946 | 0.044 | 1.06 | 2010A&A...513A....6O |
| HD110942 | 191.4957 | -43.0894 | sdB | 10.50 | -0.19 | 0.04 | 0.11 | 3.942 | 0.052 | 1.28 | TW |
| FBS2347+385 | 357.4485 | 38.7449 | sdB+WD | 11.72 | -0.19 | 0.06 | 0.11 | 3.939 | 0.040 | 0.94 | 2012MNRAS.427.2180N |
| TYC1773-399-1 | 34.4258 | 28.0582 | sdOB | 11.76 | -0.38 | 0.08 | 0.07 | 3.870 | 0.062 | 1.20 | 2021ApJS..256...28L |
| CiMelotte20488 | 50.4155 | 47.4552 | sdB+dM | 11.56 | -0.11 | 0.10 | 0.20 | 3.865 | 0.039 | 1.12 | 2012MNRAS.427.2180N |
| TYC5737-1693-1 | 297.0902 | -14.1343 | sdB | 11.62 | -0.16 | 0.13 | 0.17 | 3.805 | 0.044 | 0.98 | TW |
| [CW83]0825+15 | 127.1368 | 14.8673 | He-sdOB | 11.62 | -0.48 | 0.02 | 0.02 | 3.773 | 0.064 | 1.18 | 2012MNRAS.427.2180N |
| UCAC4 354-057819 | 153.4143 | -19.36551 | sdB | 12.57 | -0.36 | 0.02 | 0.06 | 3.769 | 0.068 | 0.95 | TW |
| TYC4470-864-1 | 324.5732 | 72.1856 | sdB | 11.33 | -0.29 | 0.05 | 0.05 | 3.765 | 0.038 | 1.12 | 2021A&A...650A.205V |
| Feige110 | 349.9933 | -5.1656 | sdOB | 11.77 | -0.49 | 0.03 | 0.02 | 3.760 | 0.062 | 1.14 | 2014A&A...566A....3R |
| UCAC4 642-091572 | 304.6748 | 38.3053 | sdO | 11.73 | -0.50 | 0.04 | 0.03 | 3.749 | 0.046 | 1.11 | TW |

| Name | RAJ2000 | DECJ2000 | Class | G | BP-RP | Stilism E(B - V) | E(44-55) | Parallax | Parallax error | RUWE | Reference |
|---|---|---|---|---|---|---|---|---|---|---|---|
| TYC467-3836-1 | 287.5225 | 2.2844 | sdB | 11.69 | -0.06 | 0.14 | 0.22 | 3.737 | 0.030 | 0.85 | 1986SAAOC..10...27K |
| PG0909+276 | 138.2153 | 27.3421 | sdOB | 12.18 | -0.44 | 0.02 | 0.04 | 3.624 | 0.079 | 0.70 | 2013A&A...557A.122G |
| TYC8147-498-1 | 120.7103 | -52.2919 | sdB | 11.87 | -0.42 | 0.04 | 0.03 | 3.609 | 0.045 | 1.03 | TW |
| TYC7709-376-1 | 155.8410 | -37.6167 | sdB+dM | 11.69 | -0.35 | 0.04 | 0.03 | 3.598 | 0.046 | 1.35 | 2013A&A...553A..18K |
| PG1352-023 | 208.7694 | -2.5061 | sdOB | 12.04 | -0.47 | 0.02 | 0.03 | 3.594 | 0.073 | 1.07 | 2021ApJS..256...28L |
| GD1068 | 30.0552 | -17.4790 | sdB+dM | 12.02 | -0.36 | 0.03 | 0.04 | 3.563 | 0.071 | 1.13 | 2016MNRAS.459.4343K |
| HD345431 | 300.1797 | 23.2672 | sdOB | 11.96 | -0.46 | 0.03 | 0.04 | 3.545 | 0.059 | 1.05 | 2021ApJS..256...28L |
| ATO J029.0051+40.0561 | 29.0052 | 40.0559 | sdB | 11.43 | -0.35 | 0.04 | 0.02 | 3.500 | 0.082 | 1.26 | 2023ApJ...942..109L |
| BD+482721 | 278.5384 | 48.4614 | sdB | 10.66 | -0.30 | 0.02 | 0.06 | 3.480 | 0.033 | 1.17 | 2013A&A...557A.122G |
| LSIV-132 | 270.5775 | -13.1749 | sdO | 11.70 | -0.02 | 0.27 | 0.32 | 3.476 | 0.042 | 0.92 | TW |
| CD-229142 | 177.6619 | -23.3430 | sdO | 11.75 | -0.50 | 0.02 | 0.00 | 3.461 | 0.062 | 1.09 | 2010Ap&SS.329..133R |
| EC14248-2647 | 216.9475 | -27.0193 | sdOB | 11.96 | -0.39 | 0.05 | 0.05 | 3.442 | 0.066 | 1.21 | 2012MNRAS.427.2180N |
| TYC5917-726-1 | 85.7715 | -16.4827 | sdB | 12.48 | -0.36 | 0.03 | 0.03 | 3.436 | 0.044 | 0.74 | TW |
| TYC8827-750-1 | 343.6857 | -55.2518 | sdB | 12.14 | -0.45 | 0.02 | 0.01 | 3.431 | 0.050 | 0.98 | 2012MNRAS.427.2180N |
| AG+81266 | 140.3296 | 81.7241 | sdO | 11.86 | -0.54 | 0.01 | 0.00 | 3.397 | 0.054 | 1.08 | 2018A&A...609A..89L |
| CD-3914181 | 319.0115 | -38.4991 | sdB | 10.95 | -0.32 | 0.01 | 0.00 | 3.328 | 0.060 | 1.14 | 2015MNRAS.453.1879K |
| UCAC4 604-016003 | 74.5074 | 30.6000 | sdB | 13.00 | 0.03 | 0.22 | 0.29 | 3.306 | 0.030 | 1.00 | TW |
| UCAC4157-007263 | 95.0479 | -58.6514 | sdO | 11.84 | -0.50 | 0.02 | 0.01 | 3.254 | 0.051 | 1.21 | TW |
| TYC9399-770-1 | 135.1694 | -77.6159 | sdB | 12.31 | -0.21 | 0.14 | 0.13 | 3.242 | 0.036 | 1.14 | TW |
| TYC2405-1118-1 | 86.1897 | 30.6481 | sdB | 12.41 | -0.24 | 0.15 | 0.13 | 3.224 | 0.053 | 0.93 | 2021ApJS..256...28L |
| HD76431 | 134.0464 | 1.6770 | He-sdO | 9.17 | -0.44 | 0.03 | 0.03 | 3.217 | 0.060 | 1.01 | 2021ApJS..256...28L |
| TYC999-2458-1 | 261.1893 | 11.5404 | sdO+WD | 12.57 | -0.29 | 0.10 | 0.09 | 3.215 | 0.061 | 1.02 | 2017A&A...600A..50G |
| CD-399849 | 231.5294 | -39.4888 | sdB | 12.07 | -0.39 | 0.06 | 0.04 | 3.212 | 0.079 | 1.20 | TW |
| HD350426 | 295.8800 | 18.4094 | sdO | 12.02 | -0.48 | 0.04 | 0.03 | 3.200 | 0.088 | 1.12 | 2018A&A...609A..89L |
| PG0133+114 | 24.1092 | 11.6588 | sdB+WD | 12.28 | -0.34 | 0.04 | 0.07 | 3.196 | 0.099 | 0.83 | 2021ApJS..256...28L |
| EC02200-2338 | 35.5828 | -23.4156 | sdB+WD | 11.99 | -0.40 | 0.02 | 0.02 | 3.162 | 0.061 | 1.07 | 2016MNRAS.459.4343K |
| PG0057+155 | 14.9863 | 15.7372 | sdOB | 12.04 | -0.43 | 0.06 | 0.04 | 3.152 | 0.062 | 1.03 | 2005A&A...430..223L |
| CPD-201123 | 91.5558 | -20.3521 | He-sdB+WD | 12.04 | -0.33 | 0.01 | 0.05 | 3.128 | 0.055 | 1.34 | 2020MNRAS.497...67L |
| PG1233+427/Feige65 | 188.9631 | 42.3776 | sdB | 11.97 | -0.41 | 0.02 | 0.01 | 3.087 | 0.054 | 1.17 | 2021ApJS..256...28L |
| LSII+2221 | 301.2396 | 22.3444 | sdO | 12.49 | -0.56 | 0.03 | 0.02 | 3.084 | 0.065 | 0.68 | 1981MNRAS.197.241W |
| GALEXJ063952.0+515658 | 99.9672 | 51.9502 | sdB | 11.94 | -0.38 | 0.04 | 0.05 | 3.084 | 0.055 | 1.05 | 2012MNRAS.427.2180N |
| HZ1 | 72.5563 | 17.7018 | He-sdO | 12.57 | -0.08 | 0.28 | 0.26 | 3.073 | 0.042 | 0.74 | 2007A&A...462..269S |
| [CW83]1419-09 | 215.6680 | -9.2896 | sdO+WD | 12.02 | -0.46 | 0.05 | 0.04 | 2.992 | 0.073 | 1.16 | 2022A&A...661A.113G |
| Feige11 | 16.0904 | 4.2268 | sdB+WD | 12.03 | -0.39 | 0.03 | 0.02 | 2.989 | 0.057 | 0.86 | 2008A&A...477L..13G |
| PG1452+198 | 223.6659 | 19.6168 | sdB+WD | 12.43 | -0.39 | 0.03 | 0.05 | 2.986 | 0.060 | 0.70 | 2021ApJS..256...28L |
| BD+182647 | 190.4658 | 17.5220 | sdO | 11.75 | -0.53 | 0.03 | 0.02 | 2.983 | 0.087 | 1.37 | 2018A&A...609A..89L |
| UCAC4282-094834 | 250.8441 | -33.7016 | sdB | 13.09 | -0.02 | 0.26 | 0.27 | 2.977 | 0.029 | 0.98 | TW |
| LAMOSTJ092239.81+270225.2 | 140.6661 | 27.0403 | sdOB+WD | 12.59 | -0.47 | 0.02 | 0.01 | 2.973 | 0.087 | 0.70 | 2021ApJS..256...28L |
| UCAC4 575-030949 | 99.1572 | 24.8570 | sdB | 12.84 | -0.31 | 0.02 | 0.00 | 2.951 | 0.060 | 0.96 | TW |
| EC15103-1557 | 228.2930 | -16.1392 | sdOB | 12.81 | -0.37 | 0.08 | 0.05 | 2.941 | 0.049 | 0.83 | 2017OAst...26..164G |
| GSC00141-01628 | 96.5928 | 4.0733 | sdB+WD | 12.03 | -0.39 | 0.03 | 0.03 | 2.928 | 0.056 | 0.92 | 2021ApJS..256...28L |
| TYC4563-2614-1 | 231.7208 | 79.6919 | sdOB | 11.64 | -0.49 | 0.03 | 0.01 | 2.927 | 0.049 | 1.31 | TW |
| TYC4824-1079-1 | 110.3765 | -4.1699 | sdB | 12.41 | -0.39 | 0.03 | 0.02 | 2.920 | 0.065 | 0.76 | TW |
| 584281747305262848O | 185.2214 | -71.4384 | sdB | 13.05 | 0.16 | 0.29 | 0.37 | 2.907 | 0.020 | 1.04 | TW |
| 2MASSJ19110167+8716038 | 287.7579 | 87.2677 | sdB | 12.71 | -0.35 | 0.02 | 0.01 | 2.885 | 0.042 | 0.78 | TW |
| UCAC4 278-224220 | 302.4073 | -34.5254 | sdOB | 12.48 | -0.38 | 0.07 | 0.07 | 2.877 | 0.052 | 1.02 | 2012MNRAS.427.2180N |
| TYC4407-187-1 | 192.7039 | 74.6621 | sdB | 11.55 | -0.41 | 0.02 | 0.01 | 2.876 | 0.042 | 1.13 | TW |
| TYC5977-517-1 | 109.9191 | -21.8895 | sdB+dM/BD | 12.13 | -0.36 | 0.03 | 0.01 | 2.869 | 0.056 | 0.91 | 2023A&A...673A..90S |
| CD-3011223 | 212.8173 | -30.8844 | sdB+WD | 12.30 | -0.37 | 0.04 | 0.03 | 2.865 | 0.066 | 0.90 | 2013A&A...554A..54G |
| HD26969/AADor | 82.9180 | -69.8837 | sdOB+dM/BD | 11.10 | -0.48 | 0.03 | 0.03 | 2.864 | 0.049 | 0.84 | 2011A&A...531L...7K |
| LSI+63198 | 52.2739 | 64.0783 | sdOB | 12.71 | -0.15 | 0.12 | 0.06 | 2.856 | 0.032 | 0.81 | 2022SchneiderPhDT |
| SBSS1709+535 | 257.5551 | 53.4461 | sdB | 12.59 | -0.38 | 0.02 | 0.01 | 2.852 | 0.045 | 0.82 | 2021ApJS..256...28L |
| UCAC4 686-011641 | 29.5873 | 47.0618 | sdB | 12.37 | -0.31 | 0.09 | 0.03 | 2.806 | 0.067 | 0.99 | TW |
| CD-403927 | 122.6323 | -40.5466 | sdO | 12.18 | -0.51 | 0.02 | 0.01 | 2.801 | 0.056 | 0.94 | 1988SAAOC..12....1K |

| Name | RAJ2000 | DECJ2000 | Class | G | BP-RP | Stilism E(B - V) | E(44-55) | Parallax | Parallax error | RUWE | Reference |
|---|---|---|---|---|---|---|---|---|---|---|---|
| LSIV+0021 | 307.8263 | 1.0904 | sdOB | 12.39 | -0.44 | 0.05 | 0.04 | 2.784 | 0.067 | 0.62 | 1991A&A...250..370B |
| PG1519+640 | 230.1309 | 63.8691 | sdB+dM/WD | 12.38 | -0.42 | 0.02 | 0.02 | 2.759 | 0.050 | 0.80 | 2013A&A...557A.122G |
| [CW83]0711+22 | 108.6244 | 22.2831 | He-sdO | 11.59 | -0.49 | 0.02 | 0.02 | 2.751 | 0.049 | 0.97 | 2021ApJS..256...28L |
| UCAC4 350-080079 | 243.6771 | -20.0086 | sdB | 12.97 | -0.01 | 0.28 | 0.31 | 2.736 | 0.038 | 0.84 | TW |
| HZ44 | 200.8966 | 36.1332 | He-sdOB | 11.61 | -0.50 | 0.02 | 0.02 | 2.735 | 0.063 | 1.33 | 2021ApJS..256...28L |
| TYC9516-559-1 | 197.9130 | -88.0869 | sdOB | 12.12 | -0.25 | 0.09 | 0.15 | 2.688 | 0.038 | 1.02 | TW |
| FBS0102+362 | 16.2035 | 36.4618 | sdOB | 12.38 | -0.43 | 0.03 | 0.03 | 2.684 | 0.067 | 0.72 | 2021ApJS..256...28L |
| PG0011+283 | 3.5927 | 28.6154 | sdB | 12.63 | -0.35 | 0.05 | 0.03 | 2.661 | 0.060 | 0.68 | 2021ApJS..256...28L |
| [CW83]0832-01 | 128.8495 | -1.9313 | He-sdO | 11.37 | -0.48 | 0.02 | 0.02 | 2.653 | 0.075 | 1.19 | 2009PhDT.......273H |
| UCAC4169-132984 | 235.3229 | -56.2659 | sdOB | 13.30 | -0.16 | 0.18 | 0.19 | 2.648 | 0.028 | 0.88 | TW |
| HD269665 | 82.7416 | -68.7529 | sdOB | 11.13 | -0.46 | 0.03 | 0.03 | 2.647 | 0.061 | 1.12 | 1989AJ.....97..881H |
| UCAC4 228-097826 | 235.3455 | -44.4066 | sdOB | 13.12 | -0.33 | 0.08 | 0.10 | 2.637 | 0.043 | 1.00 | 2021ApJS..256...28L |
| LAMOSTJ213129.05+195157.0 | 322.8711 | 19.8658 | sdO | 13.14 | -0.43 | 0.05 | 0.05 | 2.636 | 0.045 | 0.93 | 1984A&A...130..119H |
| CD-33417 | 17.1116 | -32.7199 | sdB | 12.21 | -0.40 | 0.01 | 0.00 | 2.634 | 0.070 | 1.04 | 2021ApJS..256...28L |
| TYC3281-978-1 | 31.2653 | 46.2643 | sdB | 11.50 | -0.26 | 0.08 | 0.11 | 2.630 | 0.046 | 1.09 | 2021ApJS..256...28L |
| TYC7489-686-1 | 331.4668 | -31.6844 | sdB+dM/BD | 12.33 | -0.35 | 0.01 | 0.04 | 2.630 | 0.056 | 0.74 | 2012MNRAS.427.2180N |
| TYC6343-1675-1 | 311.4515 | 21.6006 | sdOB | 12.58 | -0.44 | 0.03 | 0.04 | 2.621 | 0.063 | 0.65 | TW |
| Balloon90100001 | 348.8396 | 29.0837 | sdBV | 12.08 | -0.32 | 0.05 | 0.08 | 2.620 | 0.046 | 0.86 | 2008ASPC..392..301O |
| UCAC4 436-075435 | 272.0031 | -2.9518 | sdOB | 13.95 | 0.45 | 0.19 | 0.60 | 2.619 | 0.020 | 1.06 | TW |
| UCAC4 314-235772 | 290.9293 | -27.2806 | sdB | 13.16 | -0.36 | 0.09 | 0.09 | 2.604 | 0.060 | 1.02 | TW |
| UCAC4 219-125136 | 250.4592 | -46.3492 | sdB | 13.10 | -0.10 | 0.18 | 0.19 | 2.594 | 0.034 | 1.00 | TW |
| CPD-5510148 | 0.8891 | -54.7170 | sdB | 11.93 | -0.41 | 0.01 | 0.00 | 2.582 | 0.048 | 1.12 | TW |
| LSIV-0617 | 280.4612 | -6.4166 | sdB | 11.58 | -0.24 | 0.17 | 0.16 | 2.571 | 0.048 | 0.91 | TW |
| TYC3133-2416-1 | 288.8154 | 43.6742 | sdB | 12.44 | -0.39 | 0.02 | 0.02 | 2.566 | 0.047 | 0.73 | 2021ApJS..256...28L |
| Feige108 | 349.0517 | -1.8432 | sdB+WD | 12.98 | -0.39 | 0.04 | 0.07 | 2.555 | 0.076 | 1.01 | 1994ApJ...432..351S |
| [L92b]MarkA | 310.9969 | 46.0064 | sdO | 13.22 | -0.41 | 0.06 | 0.06 | 2.553 | 0.045 | 0.93 | 2017A&A...600A..50G |
| TYC770-941-1 | 108.0100 | 11.5590 | sdB | 12.44 | -0.39 | 0.02 | 0.02 | 2.540 | 0.071 | 0.77 | 2021ApJS..256...28L |
| UCAC4 280-113383 | 256.9890 | -34.1135 | sdB | 13.09 | 0.04 | 0.26 | 0.31 | 2.533 | 0.024 | 0.80 | TW |
| TYC2701-1209-1 | 315.8853 | 30.5938 | sdOB | 12.92 | -0.36 | 0.04 | 0.08 | 2.529 | 0.040 | 0.81 | 2021ApJS..256...28L |
| NGC 1360/CPD-26 389 | 53.3110 | -25.8715 | O(H) | 11.25 | -0.56 | 0.01 | 0.00 | 2.520 | 0.073 | 1.64 | 1977MNRAS.178..409M |
| CPD-56464 | 42.5900 | -56.2146 | sdB | 11.94 | -0.43 | 0.01 | 0.00 | 2.502 | 0.043 | 1.08 | 1991A&A...242..175V |
| PG0215+183 | 34.5658 | 18.5272 | sdB | 13.42 | -0.17 | 0.08 | 0.15 | 2.491 | 0.063 | 1.18 | 1986ApJS...61..305G |
| TYC3556-3568-1 | 294.6359 | 46.0664 | sdB+dM | 12.11 | -0.34 | 0.02 | 0.05 | 2.476 | 0.039 | 0.92 | 2010MNRAS.408L..51O |
| TYC8293-1755-1 | 226.7167 | -46.7156 | He-sdO | 12.12 | -0.37 | 0.05 | 0.12 | 2.476 | 0.076 | 0.99 | 2021ApJS..256...28L |
| LB1766 | 74.8284 | -53.8817 | He-sdB | 12.28 | -0.48 | 0.01 | 0.01 | 2.469 | 0.045 | 0.81 | 2010MNRAS.409..582N |
| GALEXJ042034.8+012041 | 65.1453 | 1.3447 | He-sdO | 12.30 | -0.44 | 0.06 | 0.05 | 2.466 | 0.079 | 0.81 | 2012MNRAS.427.2180N |
| UCAC4 190-121347 | 239.4466 | -52.0704 | sdB | 13.27 | -0.21 | 0.16 | 0.12 | 2.465 | 0.032 | 0.95 | TW |
| PG 1604+504 | 241.3922 | 50.3145 | sdOB | 12.93 | -0.44 | 0.02 | 0.00 | 2.458 | 0.047 | 0.88 | 1986ApJS...61..305G |
| UCAC4 335-187276 | 281.6388 | -23.0642 | O(H) | 13.77 | 0.21 | 0.25 | 0.45 | 2.450 | 0.028 | 1.01 | TW |
| PN A66 36/EC 13379-1937 | 205.1724 | -19.8820 | sdB | 11.49 | -0.47 | 0.06 | 0.04 | 2.450 | 0.070 | 1.26 | 2015A&A...583A.131W |
| 2200231588777454208 | 337.4641 | 58.9592 | sdB | 14.40 | -0.01 | 0.25 | 0.25 | 2.449 | 0.019 | 0.99 | ZieglerPhDT |
| UCAC4 364-079928 | 256.2398 | -17.3841 | sdOB | 13.97 | 0.02 | 0.29 | 0.34 | 2.447 | 0.025 | 0.90 | TW |
| GD1099 | 28.2118 | -16.5919 | sdB | 13.48 | -0.38 | 0.04 | 0.05 | 2.441 | 0.056 | 1.03 | 1989SAAOC..13...69K |
| CD-3019716 | 358.1505 | -30.1693 | sdO+WD | 12.05 | -0.51 | 0.01 | 0.02 | 2.427 | 0.070 | 0.98 | 2000AJ....119..241L |
| RL105 | 67.8011 | 42.9861 | sdB | 14.54 | 0.04 | 0.14 | 0.28 | 2.424 | 0.085 | 2.64 | 2021ApJS..256...28L |
| KPD0005+5106 | 2.0759 | 51.3879 | O(He) | 13.27 | -0.48 | 0.08 | 0.05 | 2.423 | 0.043 | 1.04 | 2015A&A...583A.131W |
| PG1538+401 | 235.1627 | 39.9302 | sdB | 13.20 | -0.46 | 0.02 | 0.03 | 2.423 | 0.038 | 1.00 | 2021ApJS..256...28L |
| V*EQPsc | 353.6443 | -1.3271 | sdBV+dM | 13.02 | -0.33 | 0.03 | 0.03 | 2.419 | 0.059 | 0.99 | 2019MNRAS.489.1556B |
| UCAC4 332-167523 | 278.3422 | -23.7178 | sdB | 12.81 | 0.01 | 0.25 | 0.27 | 2.408 | 0.037 | 0.81 | TW |
| TYC8532-976-1 | 87.1888 | -58.2900 | sdOB | 13.05 | -0.39 | 0.02 | 0.06 | 2.400 | 0.063 | 1.25 | 2013A&A...557A.122G |
| GALEXJ19498-2806 | 297.4519 | -28.1133 | sdB | 12.99 | -0.35 | 0.10 | 0.07 | 2.399 | 0.052 | 1.00 | 2017A&A...600A..50G |
| LSIV-14116 | 314.4120 | -14.4295 | He-sdO | 12.95 | -0.44 | 0.03 | 0.04 | 2.378 | 0.069 | 0.85 | TW |
| TYC5285-201-1 | 39.1784 | -8.8578 | sdB | 12.54 | -0.40 | 0.02 | 0.01 | 2.372 | 0.070 | 0.88 | TW |
| FBS1715+424 | 259.3283 | 42.4358 | sdO | 12.45 | -0.53 | 0.03 | 0.00 | 2.369 | 0.050 | 0.63 | 2021ApJS..256...28L |

| Name | RAJ2000 | DECJ2000 | Class | G | BP-RP | Stilism E(B - V) | E(44-55) | Parallax | Parallax error | RUWE | Reference |
|---|---|---|---|---|---|---|---|---|---|---|---|
| LAMOSTJ221002.71+250358.2 | 332.5114 | 25.0661 | sdB | 12.81 | -0.23 | 0.05 | 0.07 | 2.368 | 0.043 | 0.87 | 2021ApJS..256...28L |
| FBS1133+754 | 174.1395 | 75.1149 | sdOB | 13.15 | -0.39 | 0.03 | 0.06 | 2.365 | 0.036 | 0.88 | 2012MNRAS.427.2180N |
| KUV16256+4034 | 246.8188 | 40.4579 | sdB+dM/WD | 12.54 | -0.38 | 0.02 | 0.00 | 2.365 | 0.044 | 0.90 | 2021ApJS..256...28L |
| [CW83]1735+22 | 264.3600 | 22.1494 | sdO+WD | 11.80 | -0.45 | 0.04 | 0.04 | 2.348 | 0.055 | 1.34 | 2010A&A...519A..25G |
| 6849135629120266624 | 304.3944 | -25.1402 | sdB | 12.89 | -0.30 | 0.09 | 0.08 | 2.347 | 0.055 | 0.81 | TW |
| TYC497-63-1 | 305.8858 | 1.6057 | sdB | 12.87 | -0.32 | 0.06 | 0.08 | 2.347 | 0.056 | 0.70 | 2012MNRAS.427.2180N |
| PG1610+529 | 242.8389 | 52.7682 | sdB | 12.75 | -0.39 | 0.06 | 0.08 | 2.346 | 0.047 | 0.91 | 2021ApJS..256...28L |
| CD-249052 | 156.4620 | -24.8888 | He-sdO | 11.70 | -0.48 | 0.03 | 0.03 | 2.341 | 0.063 | 1.32 | 2009PhDT.......273H |
| UCAC4 635-004580 | 20.9384 | 36.9631 | sdB | 13.12 | -0.39 | 0.04 | 0.04 | 2.340 | 0.050 | 0.97 | TW |
| EC21027-3301 | 316.4484 | -32.8312 | sdO | 13.44 | -0.46 | 0.07 | 0.06 | 2.339 | 0.054 | 1.00 | 2015MNRAS.453.1879K |
| EC20106-5248 | 303.6082 | -52.6567 | sdB | 12.54 | -0.36 | 0.03 | 0.02 | 2.337 | 0.177 | 2.68 | 2013A&A...557A.122G |
| Feige14 | 27.0154 | -5.9294 | sdB | 12.74 | -0.38 | 0.03 | 0.01 | 2.327 | 0.058 | 0.85 | 1958ApJ...128..267F |
| JL82 | 324.0057 | -72.8076 | sdB+dM | 12.34 | -0.34 | 0.03 | 0.05 | 2.316 | 0.045 | 0.77 | 2010A&A...519A..25G |
| Feige36 | 166.1317 | 24.6619 | sdB+WD | 12.72 | -0.39 | 0.02 | 0.03 | 2.312 | 0.056 | 0.95 | 1999ASPC..169..546E |
| UCAC4 507-015874 | 88.2498 | 11.3128 | sdOB | 13.32 | -0.16 | 0.24 | 0.22 | 2.284 | 0.032 | 0.94 | TW |
| TYC7691-3990-1 | 140.5535 | -39.0237 | sdB | 12.73 | -0.36 | 0.04 | 0.03 | 2.283 | 0.040 | 0.77 | TW |
| JL163 | 2.6385 | -50.2567 | sdB | 12.92 | -0.41 | 0.01 | 0.00 | 2.276 | 0.041 | 0.73 | 1989SAAOC..13...69K |
| GALEX J191509.0-290311 | 288.7882 | -29.0527 | sdB | 13.22 | -0.29 | 0.09 | 0.11 | 2.275 | 0.044 | 1.04 | 2017A&A...600A..50G |
| HZ3 | 58.3806 | 10.7511 | He-sdO | 12.80 | -0.27 | 0.20 | 0.15 | 2.264 | 0.072 | 0.98 | 2021ApJS..256...28L |
| TYC5859-1253-1 | 32.4138 | -18.1739 | sdB | 13.12 | -0.43 | 0.01 | 0.01 | 2.256 | 0.047 | 1.02 | 2022schneiderPhDT |
| Feige95 | 217.1232 | 21.1006 | sdB | 13.21 | -0.42 | 0.02 | 0.04 | 2.247 | 0.057 | 1.34 | 2021ApJS..256...28L |
| PG0314+146 | 49.4084 | 14.7732 | He-sdO | 12.50 | -0.19 | 0.18 | 0.19 | 2.245 | 0.065 | 1.03 | 2012MNRAS.427.2180N |
| PG1232-136 | 188.8279 | -13.9191 | sdB+dM/WD | 13.23 | -0.37 | 0.02 | 0.03 | 2.245 | 0.042 | 0.92 | 2015A&A...576A..44K |
| EVR-CB-001 | 132.0645 | -74.3151 | sdB+WD | 12.57 | -0.16 | 0.10 | 0.09 | 2.243 | 0.039 | 0.94 | 2019ApJ...883...51R |
| 2MASSJ09112370-2415462 | 137.8488 | -24.2629 | sdOB | 13.57 | -0.27 | 0.07 | 0.12 | 2.238 | 0.032 | 1.03 | TW |
| TYC4651-1475-1 | 248.0599 | 85.2329 | sdB | 12.32 | -0.34 | 0.12 | 0.09 | 2.231 | 0.044 | 0.85 | 2012MNRAS.427.2180N |
| LAMOSTJ180933.32+223059.9 | 272.3890 | 22.5167 | sdB | 12.92 | -0.27 | 0.07 | 0.10 | 2.230 | 0.035 | 0.84 | 2021ApJS..256...28L |
| HD265435 | 103.3512 | 33.0595 | sdOB + WD | 12.09 | -0.43 | 0.03 | 0.04 | 2.216 | 0.069 | 1.06 | 2021ApJS..256...28L |
| LSIV+092 | 263.9314 | 9.1436 | sdB+dM/BD | 12.68 | -0.20 | 0.12 | 0.14 | 2.214 | 0.050 | 0.77 | 2017A&A...600A..50G |
| PG2337+070 | 355.0197 | 7.2858 | sdB | 13.45 | -0.15 | 0.08 | 0.18 | 2.210 | 0.038 | 1.08 | 2021ApJS..256...28L |
| 2MASSJ02065617+1438585 | 31.7341 | 14.6495 | sdB | 13.34 | -0.34 | 0.07 | 0.08 | 2.210 | 0.054 | 1.05 | 2012MNRAS.427.2180N |
| GALEXJ07015-6717 | 105.3926 | -67.2946 | sdOB | 13.13 | -0.39 | 0.04 | 0.05 | 2.198 | 0.032 | 1.07 | 2017A&A...600A..50G |
| TYC9249-242-1 | 193.0616 | -72.9138 | sdB | 12.97 | -0.19 | 0.16 | 0.14 | 2.192 | 0.034 | 1.00 | TW |
| Ton788 | 228.6350 | 24.1780 | sdB+WD | 13.18 | -0.36 | 0.05 | 0.06 | 2.190 | 0.041 | 1.10 | 2021ApJS..256...28L |
| KUV20417+7604 | 310.1953 | 76.2444 | sdB | 12.73 | -0.20 | 0.17 | 0.14 | 2.187 | 0.036 | 0.99 | 1987AJ.....94.1271W |
| 5881540795158796416 | 224.2913 | -55.4395 | sdB | 14.06 | -0.13 | 0.15 | 0.21 | 2.180 | 0.027 | 0.99 | TW |
| PG2349+002 | 357.9717 | 0.4715 | sdB | 13.26 | -0.34 | 0.03 | 0.07 | 2.176 | 0.043 | 1.08 | 2013A&A...557A.122G |
| TYC1969-78-1 | 155.2731 | 22.9907 | sdO+dM | 12.66 | -0.42 | 0.02 | 0.00 | 2.176 | 0.069 | 0.63 | TW |
| UCAC4 095-055105 | 219.0558 | -71.1940 | sdOB | 13.33 | -0.21 | 0.12 | 0.12 | 2.171 | 0.038 | 1.00 | TW |
| PG1710+490 | 258.0782 | 48.9766 | sdBV | 12.36 | -0.40 | 0.01 | 0.00 | 2.167 | 0.041 | 0.77 | 2013A&A...557A.122G |
| TYC4542-482-1 | 154.5052 | 75.2244 | sdB+WD | 13.28 | -0.45 | 0.03 | 0.03 | 2.166 | 0.039 | 0.83 | 2022ApJ...928...20B |
| US719 | 143.4494 | 46.0750 | sdOB | 13.39 | -0.33 | 0.07 | 0.05 | 2.166 | 0.039 | 0.66 | 2017A&A...600A..50G |
| PB7352 | 343.9300 | -6.9944 | sdO | 13.57 | -0.45 | 0.03 | 0.10 | 2.153 | 0.062 | 0.91 | 2015A&A...576A..44K |
| Feige91 | 212.1341 | 59.6738 | sdB+dM/WD | 12.27 | -0.41 | 0.02 | 0.00 | 2.143 | 0.069 | 0.97 | 1974ApJS...28..157G |
| 5917224379853806712 | 257.1978 | -56.6804 | sdB | 13.40 | -0.45 | 0.02 | 0.03 | 2.141 | 0.033 | 1.04 | TW |
| EC11575-1845 | 180.0235 | -19.0344 | sdO+dM | 14.01 | -0.18 | 0.12 | 0.12 | 2.138 | 0.030 | 2.32 | 2015ApJ...808..179D |
| PG1207-033 | 182.4004 | -3.5521 | sdOB | 13.09 | -0.26 | 0.04 | 0.16 | 2.138 | 0.067 | 0.99 | 2005A&A...430..223L |
| CD-48106 | 7.9237 | -47.4223 | sdBV | 13.33 | -0.45 | 0.03 | 0.03 | 2.134 | 0.058 | 0.74 | 1984A&A...130..119H |
| PG0918+029 | 140.3674 | 2.7672 | sdB+WD | 12.36 | -0.40 | 0.01 | 0.00 | 2.134 | 0.049 | 1.04 | 2021ApJS..256...28L |
| FBS2154+329 | 329.2756 | 33.1365 | sdOB | 13.28 | -0.45 | 0.03 | 0.03 | 2.133 | 0.059 | 1.01 | 2021ApJS..256...28L |
| LAMOSTJ064618.36+292013.2 | 101.5765 | 29.3370 | sdO | 13.57 | -0.33 | 0.07 | 0.07 | 2.123 | 0.035 | 1.04 | 2008AJ....136..946M |
| UCAC4 435-106149 | 298.7121 | -3.1255 | He-sdOB | 13.22 | 0.02 | 0.22 | 0.36 | 2.118 | 0.050 | 0.95 | 2021ApJS..256...28L |
| UCAC4 174-038075 | 146.4025 | -55.2652 | sdB | 12.82 | -0.36 | 0.04 | 0.02 | 2.114 | 0.021 | 0.82 | TW |
| UCAC4 198-195239 | 349.9556 | -50.5634 | sdOB | 13.07 | -0.48 | 0.01 | 0.00 | 2.109 | 0.046 | 0.94 | TW |
|  |  |  |  |  |  |  |  | 2.103 | 0.043 |  |  |

| Name | RAJ2000 | DECJ2000 | Class | G | BP-RP | Stilism E(B - V) | E(44-55) | Parallax | Parallax error | RUWE | Reference |
|---|---|---|---|---|---|---|---|---|---|---|---|
| FBS2253+335 | 343.9925 | 33.7199 | sdB | 12.75 | -0.26 | 0.07 | 0.12 | 2.102 | 0.050 | 0.85 | 2021ApJS..256...28L |
| Feige38 | 169.2056 | 6.9925 | sdB | 12.98 | -0.37 | 0.03 | 0.06 | 2.098 | 0.071 | 1.04 | 2013A&A...557A.122G |
| TonS183 | 15.3232 | -33.7127 | sdB+WD | 12.58 | -0.41 | 0.01 | 0.01 | 2.090 | 0.059 | 0.76 | 2010A&A...519A..25G |
| PG1722+286 | 261.0498 | 28.5908 | sdB | 13.33 | -0.45 | 0.03 | 0.04 | 2.090 | 0.033 | 1.02 | 2021ApJS..256...28L |
| PG0004+133 | 1.8907 | 13.5992 | sdB | 13.05 | -0.13 | 0.05 | 0.22 | 2.090 | 0.036 | 0.93 | 2021ApJS..256...28L |
| TYC769-1149-1 | 114.5485 | 10.6264 | sdB | 11.80 | -0.45 | 0.02 | 0.03 | 2.076 | 0.071 | 1.19 | TW |
| TYC1117-348-1 | 318.6130 | 13.3685 | sdOB | 13.69 | -0.31 | 0.05 | 0.03 | 2.076 | 0.066 | 1.31 | TW |
| PHL1548 | 52.2809 | -11.7056 | sdOB | 13.26 | -0.37 | 0.03 | 0.07 | 2.074 | 0.038 | 1.13 | 2013A&A...557A.122G |
| UCAC4 178-120735 | 213.7422 | -54.4676 | sdOB | 13.48 | -0.14 | 0.20 | 0.09 | 2.073 | 0.028 | 1.06 | TW |
| BPSCS22947-0196 | 290.9424 | -47.7881 | sdB | 13.29 | -0.36 | 0.04 | 0.21 | 2.068 | 0.040 | 1.01 | 1992AJ...103..267B |
| UCAC4 540-131404 | 305.7445 | 17.8390 | sdOB | 13.02 | -0.45 | 0.03 | 0.09 | 2.066 | 0.047 | 0.87 | TW |
| HE0247-0418 | 42.5997 | -4.1040 | sdB | 13.00 | -0.39 | 0.03 | 0.03 | 2.060 | 0.060 | 0.92 | 2012MNRAS.427.2180N |
| EC21556-5552 | 329.7529 | -55.6343 | sdB+WD | 13.10 | -0.39 | 0.01 | 0.02 | 2.049 | 0.040 | 1.04 | 2011MNRAS.415.1381C |
| PG1634+061 | 249.2649 | 5.9848 | sdB | 13.82 | -0.34 | 0.06 | 0.08 | 2.048 | 0.096 | 2.78 | 2021ApJS..256...28L |
| UCAC4 285-089547 | 243.7561 | -33.1560 | sdOB | 13.46 | -0.23 | 0.28 | 0.17 | 2.047 | 0.035 | 1.05 | TW |
| UCAC4 575-018970 | 86.1719 | 24.9196 | sdB | 13.60 | -0.21 | 0.13 | 0.13 | 2.036 | 0.039 | 1.08 | TW |
| Feige46 | 174.3595 | 14.1704 | He-sdO | 13.24 | -0.48 | 0.02 | 0.02 | 2.035 | 0.060 | 1.00 | 2019ApJ...881....7L |
| Feige109 | 349.3620 | 7.8680 | sdB+WD | 13.72 | -0.28 | 0.05 | 0.09 | 2.032 | 0.043 | 1.20 | 2013A&A...557A.122G |
| GALEXJ18599-6749 | 284.9928 | -67.8287 | sdB | 13.22 | -0.35 | 0.04 | 0.05 | 2.030 | 0.034 | 1.02 | 2017A&A...600A..50G |
| LSIV+109 | 310.7603 | 10.5725 | He-sdO | 11.95 | -0.44 | 0.08 | 0.05 | 2.030 | 0.060 | 1.05 | 2018A&A...620A..36S |
| PG1619+522 | 245.1613 | 52.1025 | sdB+WD | 13.23 | -0.45 | 0.02 | 0.03 | 2.027 | 0.040 | 1.10 | 1994ApJ...432..351S |
| LSII+091 | 299.6914 | 10.0190 | sdB | 12.79 | -0.33 | 0.07 | 0.04 | 2.026 | 0.056 | 0.81 | TW |
| EC23073-6905 | 347.6483 | -68.8252 | sdB | 12.54 | -0.40 | 0.02 | 0.01 | 2.022 | 0.048 | 0.88 | 2022MöllerMScT |
| 2MASSJ21352819+4903391 | 323.8677 | 49.0609 | sdB | 13.61 | -0.30 | 0.08 | 0.03 | 2.017 | 0.027 | 1.11 | TW |
| UCAC4 202-168220 | 271.7187 | -49.6211 | sdB | 13.13 | -0.28 | 0.09 | 0.09 | 2.016 | 0.045 | 1.05 | TW |
| PG1544+488 | 236.5484 | 48.6438 | He-sdOB+He-sdOB | 12.77 | -0.44 | 0.02 | 0.05 | 2.014 | 0.046 | 0.75 | 2021ApJS..256...28L |
| GALEXJ10415+1842 | 160.3767 | 18.7027 | sdOB | 12.97 | -0.46 | 0.02 | 0.02 | 2.014 | 0.068 | 1.10 | 2021ApJS..256...28L |
| UCAC4 149-221427 | 354.7874 | -60.2195 | sdB | 13.31 | -0.37 | 0.01 | 0.00 | 2.009 | 0.033 | 1.04 | TW |
| LAMOSTJ044847.21-040016.9 | 72.1968 | -4.0049 | sdB | 13.35 | -0.39 | 0.03 | 0.05 | 2.007 | 0.049 | 1.18 | 2021ApJS..256...28L |
| 2MASSJ15292631+7011543 | 232.3599 | 70.1983 | sdB+dM/BD | 12.44 | -0.36 | 0.02 | 0.00 | 2.006 | 0.048 | 0.78 | 2022ApJ...928...20B |
| GALEXJ19111-1406 | 287.7885 | -14.1149 | He-sdO | 11.88 | -0.27 | 0.17 | 0.14 | 2.005 | 0.054 | 1.00 | 2021MNRAS.501.623J |
| 60284367856423851 52 | 254.9810 | -31.2995 | sdB | 14.01 | -0.21 | 0.16 | 0.11 | 2.005 | 0.035 | 0.97 | TW |
| UCAC4 299-079624 | 229.5817 | -30.2446 | sdB | 13.72 | 0.00 | 0.20 | 0.31 | 2.004 | 0.024 | 0.97 | TW |
| TYC 6800-72-1 | 249.6178 | -24.8532 | He-sdO | 12.11 | -0.05 | 0.29 | 0.32 | 2.002 | 0.034 | 0.83 | TW |
| UCAC4488-012508 | 83.2798 | 7.5061 | sdOB | 13.79 | -0.17 | 0.18 | 0.21 | 1.999 | 0.032 | 1.07 | 2021ApJS..256...28L |
| MWP 1 | 319.2845 | 34.2077 | PG1159 | 13.02 | -0.53 | 0.25 | 0.03 | 1.991 | 0.045 | 0.95 | 1993A&A...268..561M |
| UCAC4 451-072127 | 267.4836 | 0.1100 | sdOB | 14.34 | 0.09 | 0.06 | 0.36 | 1.984 | 0.025 | 1.02 | 2021ApJS..256...28L |
| UCAC4 497-042535 | 110.6667 | 9.2318 | sdO | 13.03 | -0.52 | 0.02 | 0.01 | 1.984 | 0.067 | 0.90 | 2021ApJS..256...28L |
| UCAC4 067-027932 | 260.0518 | -76.6077 | sdB | 13.54 | -0.18 | 0.05 | 0.16 | 1.982 | 0.022 | 1.04 | 2017A&A...600A..50G |
| EC11119-2405 | 168.5918 | -24.3582 | sdB | 12.76 | -0.26 | 0.03 | 0.07 | 1.976 | 0.053 | 0.84 | 2011MNRAS.410.2095V |
| GD1110 | 349.8517 | -8.8773 | sdB | 12.92 | -0.39 | 0.05 | 0.03 | 1.967 | 0.075 | 0.81 | 2010A&A...513A...6O |
| LSIII+4850 | 337.4885 | 48.3578 | sdOB | 12.23 | -0.32 | 0.02 | 0.07 | 1.965 | 0.041 | 0.94 | TW |
| UCAC4 349-002972 | 43.2902 | -20.2426 | sdO | 12.03 | -0.49 | 0.02 | 0.01 | 1.955 | 0.083 | 0.97 | TW |
| EC12408-1427 | 190.8751 | -14.7303 | sdB+WD | 12.78 | -0.39 | 0.02 | 0.03 | 1.939 | 0.065 | 0.66 | 1997MNRAS.287..867K |

TW (This Work): Spectroscopically identified or reclassified in this work.

Table A.2: 500 pc sample of composite hot subdwarf stars from *Gaia* DR3 ad identified through IR excess in their SED fits (Sec. 5.3), sorted by increasing distance.

| Name | RAJ2000 | DECJ2000 | Class | G | BP-RP | Stilism E(B - V) | Parallax | Parallax error | RUWE | Reference |
|---|---|---|---|---|---|---|---|---|---|---|
| BD+102357 | 178.9860 | 9.8471 | sdO+A8V* | 8.81 | 0.16 | 0.01 | 5.883 | 0.035 | 1.21 | 1980A&A....85..367B |
| HD185510 | 294.9119 | -6.0639 | sdB+KOIII* | 7.80 | 1.39 | 0.11 | 5.600 | 0.028 | 0.97 | 1992AJ....103..267B |
| SB290 | 10.7432 | -38.1271 | sdB+K7V | 10.42 | -0.32 | 0.00 | 5.485 | 0.164 | 3.18 | 2013A&A...557A.122G |
| BD+341543 | 107.5323 | 34.4147 | sdB+F9V* | 10.05 | 0.33 | 0.01 | 5.143 | 0.044 | 2.05 | 2013A&A...559A..54V |
| BD-035357 | 330.1519 | -2.7408 | sdOB+G8III* | 9.06 | 1.20 | 0.07 | 4.770 | 0.039 | 2.05 | 1982MNRAS.201..901D |
| CD-243988 | 96.0465 | -24.5307 | sdOB+G6V | 10.68 | 0.28 | 0.01 | 4.406 | 0.086 | 4.93 | TW |
| GD319 | 192.5183 | 55.1004 | sdB+K3V | 12.67 | -0.42 | 0.02 | 4.039 | 0.053 | 0.72 | 2021ApJS..256...28L |
| PG1104+243 | 166.8590 | 24.0530 | sdOB+G0V | 11.20 | 0.22 | 0.02 | 3.467 | 0.070 | 2.30 | 2012A&A...548A...6V |
| CD-441028 | 48.3133 | -43.9164 | sdOB+F7V | 10.72 | 0.22 | 0.01 | 3.458 | 0.049 | 2.08 | TW |
| TYC4454-1229-1 | 300.9430 | 71.6068 | sdO+G5V | 10.40 | 0.02 | 0.04 | 3.452 | 0.092 | 3.56 | TW |
| BD+293070 | 264.5883 | 29.1466 | sdB+F8V* | 10.29 | 0.34 | 0.02 | 3.428 | 0.043 | 2.43 | 2013A&A...559A..54V |
| PG0749+658 | 118.6019 | 65.7021 | sdB+K1V | 12.06 | -0.02 | 0.03 | 3.370 | 0.132 | 4.49 | 1994ApJ...432..351S |
| TYC4890-19-1 | 135.5186 | -7.3465 | sdOB+M0[(TW)] | 12.02 | -0.35 | 0.02 | 3.335 | 0.069 | 1.20 | 2017MNRAS.466.5020H |
| CPD-73420 | 104.4036 | -73.4133 | sdB+G | 11.74 | -0.17 | 0.09 | 3.290 | 0.038 | 1.18 | 2012MNRAS.427.2180N |
| GD274 | 16.7959 | 51.1727 | sdOB+G5V | 12.08 | 0.37 | 0.09 | 3.281 | 0.105 | 4.59 | 2021ApJS..256...28L |
| TYC7831-110-1 | 220.5407 | -43.2968 | sdB+K3V | 11.78 | -0.30 | 0.06 | 3.199 | 0.144 | 2.30 | TW |
| BD-11162 | 13.0626 | -10.6629 | sdO+G0V | 11.08 | 0.01 | 0.03 | 3.063 | 0.119 | 3.69 | 1953PASP...65...48V |
| CPD-71172 | 43.3781 | -71.3756 | sdOB+F2V* | 10.61 | 0.30 | 0.03 | 3.059 | 0.021 | 0.93 | 1988A&A...205..147V |
| TYC5337-1128-1 | 83.7761 | -10.5912 | sdB+F6V | 11.57 | 0.23 | 0.03 | 2.976 | 0.055 | 2.18 | TW |
| GD1053 | 29.1329 | -13.9078 | sdOBV+K1V[(TW)] | 12.25 | -0.26 | 0.05 | 2.866 | 0.140 | 2.02 | 2010A&A...513A...6O |
| TYC 1703-394-1 | 336.9935 | 20.1063 | sdB+F5V* | 10.64 | 0.39 | 0.04 | 2.858 | 0.032 | 1.61 | 2012MNRAS.427.2180N |
| SDSSJ185859.95 | 284.7499 | 17.3032 | sdB+F8V* | 11.83 | 0.58 | 0.13 | 2.809 | 0.042 | 2.11 | 2017A&A...600A..50G |
| LSIV+065 | 285.3290 | 6.2493 | sdOB+K1V | 12.35 | -0.15 | 0.23 | 2.773 | 0.046 | 1.11 | TW |
| TYC6906-318-1 | 305.2490 | -22.8341 | sdB+G2V | 11.68 | 0.26 | 0.09 | 2.710 | 0.086 | 3.82 | 2012MNRAS.427.2180N |
| UCAC4 235-080231 | 223.6418 | -43.0661 | sdB+G9V | 12.76 | 0.08 | 0.06 | 2.699 | 0.094 | 2.98 | TW |
| CD-488608 | 209.1215 | -49.5676 | sdOB+K0V | 12.16 | -0.06 | 0.07 | 2.677 | 0.061 | 1.52 | 2012MNRAS.427.2180N |
| TYC2084-448-1 | 264.2134 | 28.1096 | sdB+F6V* | 11.45 | 0.40 | 0.03 | 2.605 | 0.022 | 1.31 | 2012MNRAS.427.2180N |
| TYC7644-1857-1 | 114.2603 | -38.9095 | sdO+G2V | 12.65 | 0.19 | 0.08 | 2.529 | 0.054 | 2.58 | TW |
| EC05053-2806 | 76.8343 | -28.0403 | sdB+G0V | 12.43 | 0.09 | 0.01 | 2.495 | 0.043 | 1.29 | 2013MNRAS.431.240O |
| TYC1337-283-1 | 98.8248 | 20.5924 | He-sdOB+A0V* | 11.40 | 0.25 | 0.02 | 2.488 | 0.414 | 6.62 | 2019ApJ...881....7L |
| UCAC4 275-013588 | 104.3408 | -35.1936 | sdOB+K1V | 13.15 | 0.13 | 0.08 | 2.446 | 0.103 | 5.55 | TW |
| 4417791888809043328 | 233.0570 | 1.0806 | sdOB+K7V | 12.77 | -0.15 | 0.05 | 2.405 | 0.123 | 2.31 | TW |
| HDE283048 | 58.3072 | 25.7560 | He-sdO+FIV* | 10.12 | 0.49 | 0.13 | 2.386 | 0.055 | 2.71 | 2023ApJ...942..109L |
| Feige87 | 205.0614 | 60.8796 | sdB+G0V | 11.61 | -0.05 | 0.02 | 2.302 | 0.097 | 3.73 | 2012ApJ...758...58B |
| TYC4499-2297-1 | 302.4654 | 3.1755 | sdB+K3V[(TW)] | 12.66 | -0.20 | 0.07 | 2.299 | 0.141 | 2.33 | 2012MNRAS.427.2180N |
| ClMelotte 22 AK IV-859 | 60.8945 | 26.0663 | sdB+F7V* | 12.27 | 0.37 | 0.20 | 2.291 | 0.049 | 2.04 | 2023ApJ...942..109L |
| PG2317+046 | 349.9807 | 4.8762 | sdO+KV | 12.79 | -0.29 | 0.05 | 2.290 | 0.065 | 0.93 | 2013A&A...557A.122G |
| GD299 | 144.5848 | 55.0969 | HesdOB+K3V | 12.01 | -0.35 | 0.02 | 2.265 | 0.056 | 1.15 | 2021ApJS..256...28L |
| UCAC4 716-006410 | 11.8897 | 53.0791 | sdB+K6V | 12.94 | -0.16 | 0.09 | 2.241 | 0.039 | 0.89 | TW |
| SB744 | 27.1839 | -26.6038 | sdOB+G1V | 12.22 | 0.03 | 0.01 | 2.221 | 0.095 | 2.60 | 2000AJ....119..241L |
| NGC 1514/HD 281679 | 62.3207 | 30.7760 | O(H)+A* | 9.28 | 0.81 | 0.29 | 2.203 | 0.019 | 1.05 | 1997PASP..109..659F |
| TYC3871-835-1 | 228.9095 | 56.8955 | sdB+G0V* | 11.37 | 0.32 | 0.02 | 2.164 | 0.036 | 1.95 | 2014ASPC..481..265V |
| EC11031-1348 | 166.4225 | -14.0734 | sdB+F6* | 11.43 | 0.44 | 0.03 | 2.156 | 0.027 | 0.98 | 2012MNRAS.427.2180N |
| TYC4401-1269-1 | 216.7852 | 72.9638 | sdB+F3V* | 11.14 | 0.19 | 0.02 | 2.105 | 0.027 | 1.51 | TW |
| EC20217-5704 | 306.4240 | -56.9140 | sdOB+K3V[(TW)] | 12.43 | -0.30 | 0.03 | 2.030 | 0.109 | 1.32 | 2013MNRAS.431.240O |
| EC14429-2701 | 176.3692 | -27.3066 | sdB+F2V* | 11.29 | 0.39 | 0.03 | 2.022 | 0.033 | 1.03 | 1997MNRAS.287..867K |
| GALEXJ02342+2342 | 38.5552 | 23.7030 | sdB+F5V* | 12.14 | 0.54 | 0.10 | 2.021 | 0.024 | 1.22 | 2017A&A...600A..50G |
| PHL 1079 | 24.6125 | 3.6608 | sdB+G7 | 13.27 | 0.11 | 0.03 | 1.924 | 0.108 | 4.67 | 2012ApJ...758...58B |

TW (This Work): Spectroscopically identified or reclassified in this work. [(TW)] Previously known hot subdwarf system, but with a newly identified cool MS companion.
* (asterix) Previously known systems manually added to the sample. Companion spectral classes are derived from our SED based effective temperature estimates using the calibration given in Cox (2000).

Table A.3: Other underluminous objects

| Name | RAJ2000 | DECJ2000 | Class | Reference |
|---|---|---|---|---|
| PB5919 | 5.8918 | 6.9465 | CV | TW |
| ATO J005.9853+42.7854 | 5.9854 | 42.7855 | CV | 2023AJ....165..148H |
| PG0038+199 | 10.3974 | 20.1546 | DO1 | 2017A&A...601A...8W |
| HD9478 | 22.7583 | -66.4967 | BV9 | 1975mcts.book.....H |
| V*BGTri | 26.1982 | 32.5499 | CV | 2008PZP.....8....4K |
| V*BOCet | 31.6636 | -2.0619 | CV | 1993PASP..105..127D |
| BD+14341 | 31.7211 | 15.2949 | CV | 1966PASP...78..279W |
| 300394067131824768 | 31.8226 | 30.0865 | WD | 2022yCat.5156....0L |
| UCAC4 693-013549 | 31.9930 | 48.5754 | CV | 2017AJ....153..204S |
| 2MASS J02150626+0155041 | 33.7760 | 1.9176 | sdA | 2023ApJ...950..141K |
| FBS0212+385 | 33.9933 | 38.7721 | DAO | 2021ApJS..256...28L |
| 1RXS J022917.1-395851 | 37.3208 | -39.9838 | CV | 2018MNRAS.473..693V |
| HS0229+8016 | 38.9930 | 80.4957 | CV | 2005A&A...443..995A |
| V*IMEri | 66.1715 | -20.1200 | CV | 2001PASP..113..764D |
| HD30112 | 71.1753 | 0.5680 | B3/5V | 1999MSS...C05....0H |
| HD31726 | 74.4362 | -14.2320 | MSB | TW |
| V*V1159Ori | 82.2481 | -3.5646 | CV | 1992A&A...259..198J |
| V*UYOriA | 83.0013 | -4.9316 | Herbig Ae | 2015MNRAS.453..976F |
| V*TWPic | 83.7107 | -58.0280 | CV | 1984BAAS...16..514S |
| V*CNOri | 88.0324 | -5.4168 | CV | TW |
| TYC1325-1524-1 | 90.4308 | 21.5494 | BHB | TW |
| HD43112 | 93.7854 | 13.8512 | B1V | 2020A&A...639A..81B |
| TYC144-2049-1 | 94.6424 | 5.8229 | MSA/BHB | TW |
| UCAC4 316-011308 | 95.9167 | -26.9641 | CV | 2020ATel14219....1S |
| EGB4 | 97.3915 | 71.0766 | CV | 1983ApJS...53..523W |
| HD47144B | 98.8503 | -36.7799 | - | - |
| V*RR Pic | 98.9003 | -62.6401 | CV | 1950PASP...62..221H |
| 3378090704289927552 | 100.6357 | 20.9336 | - | - |
| TYC6526-2311-1 | 104.6056 | -25.4155 | MSA/BHB | TW |
| ASASJ071404+7004.3 | 108.5197 | 70.0716 | CV | 2022MNRAS.510.3605I |
| TYC6545-1888-1 | 110.1178 | -27.6298 | BHB | TW |
| 3053791570751929344 | 112.0746 | -8.1706 | - | - |
| 2MASSJ07465548-0934305 | 116.7312 | -9.5752 | CV | 2008MNRAS.385.1485P |
| TYC5999-40-1 | 121.2186 | -17.0072 | MSA/BHB | TW |
| TYC6560-514-1 | 123.5778 | -24.8821 | - | - |
| V*IXVel | 123.8288 | -49.2228 | CV | 1984AJ.....89..389E |
| PG0834+501 | 129.4056 | 49.8743 | DAO.8 | 2011ApJ...743..138G |
| 586495788772380672 | 142.3875 | 7.7051 | MSA/BHB | 2022yCat.5156....0L |
| V*ERUMa | 146.8000 | 51.9025 | CV | 1993PASP..105..127D |
| TYC834-791-1 | 147.0800 | 13.4886 | MSA/BHB | TW |
| V*RWSex | 154.9858 | -8.6990 | CV | 1974ApJ...189L.131G |
| PG1034+001 | 159.2654 | -0.1385 | DO | 1983BAAS...15Q.984S |
| BD-112966 | 162.9262 | -12.4026 | BHB | TW |
| 5389717630410364160 | 165.0685 | -42.6771 | CV | TW |
| CD-416382 | 167.6733 | -42.2374 | BHB | TW |
| 5371969485515378176 | 174.5655 | -48.5888 | - | - |
| HS1136+6646 | 174.7733 | 66.5048 | DAO+KV | 2021A&A...647A.184R |
| 5377677187816453632 | 177.2208 | -46.4381 | - | - |
| 6058834949182961536 | 180.9110 | -60.3801 | sdA | TW |
| Feige55 | 181.1601 | 60.5354 | DAO | 2011ApJ...743..138G |
| V*AMCVn | 188.7278 | 37.6290 | AMCVn | 1966ApJ...144..496G |
| TYC7791-1293-1 | 201.1430 | -41.1065 | CV | 1995PASP..107..846D |
| V*UXUMa | 204.1703 | 51.9138 | CV | 1949ApJ...110..387M |
| BD+302431 | 204.6031 | 29.3651 | BHB | 2018A&A...618A..86S |
| V*V827 Cen | 206.0665 | -51.0125 | ApSiCr | 1978mcts.book.....H |
| CPD-731185 | 206.2093 | -74.1253 | BHB | TW |
| HD135485 | 228.9386 | -14.6931 | B3V | 1988mcts.book.....H |
| V*HPLib | 233.9710 | -14.2201 | AMCVn | 1994MNRAS.271..910O |

Table A.3: Other underluminous objects

| Name | RAJ2000 | DECJ2000 | Class | Reference |
|---|---|---|---|---|
| V*LXSer | 234.5003 | 18.8676 | CV | 1979IBVS.1630....1S |
| 2MASSJ16360160+5944411 | 249.0069 | 59.7447 | CV | 2022ApJ...928...20B |
| LSIV-08 1 | 249.0907 | -8.1101 | BHB | TW |
| V*V1084Her | 250.9404 | 34.0443 | CV | 2001PASP..113..764D |
| V*V341Ara | 254.4225 | -63.2111 | CV | 1995PASP..107..846D |
| ATO J256.7330-24.9840 | 256.7331 | -24.9841 | WD+dM? | TW |
| 2MASSJ17514575+3820157 | 267.9406 | 38.3378 | DA | 2023A&A...677A..29R |
| TYC4213-1610-1 | 270.3466 | 65.9479 | MSF | 1981ApJS...45..437A |
| TYC8357-3863-1 | 272.1239 | -45.7673 | - | - |
| CPD-634369 | 274.7516 | -63.3007 | CV | 1995PASP..107..846D |
| V*V603Aql | 282.2277 | 0.5841 | CV | TW |
| HD337604 | 286.4118 | 25.6505 | pre-ELM/sdB candidate | TW |
| V*MVLyr | 286.8179 | 44.0188 | CV | 1981ApJ...245..644S |
| HD231084 | 289.4615 | 13.4102 | BHB | TW |
| TYC3556-325-1 | 292.8714 | 45.9849 | CV | 2013ApJ...775...64G |
| 4318061098980872960 | 294.0047 | 14.5094 | sdA | TW |
| TYC3135-86-1 | 294.7834 | 37.5719 | pre-ELM/sdB candidate | TW |
| HD186605A | 295.9462 | 38.3223 | MSB | TW |
| 6684666865906605568 | 296.2970 | -44.9985 | DAO | 2023A&A...677A..29R |
| CD-4214462 | 296.9191 | -42.0075 | CV | 1971PASP...83..485B |
| V*ABDrac | 297.2770 | 77.7398 | CV | 1995A&AS..114..269D |
| M27 | 299.9016 | 22.7212 | DAO.6 | 2011ApJ...743..138G |
| ATO J300.8707+08.6463 | 300.8707 | 8.6464 | DAO+MS | 2023AJ....165..142K |
| HD194375 | 303.6092 | 80.5317 | MSB | TW |
| V*CMDel | 306.2372 | 17.2983 | CV | 1982A&AS...48..383V |
| HBHA4204-09 | 316.9677 | 44.0950 | CV | 1997AAHam..11....1K |
| TYC546-1349-1 | 323.5281 | 3.5709 | CV | 2021ApJS..256...28L |
| V*LSPeg | 327.9915 | 14.1147 | CV | 1986ApJ...300..779S |
| V*UUAqr | 332.2740 | -3.7717 | CV | 1986ApJ...300..779S |
| NGC 7293/PHL 287 | 337.4108 | -20.8372 | DAO.5 | 2011ApJ...743..138G |
| V*AOPsc | 343.8250 | -3.1779 | CV | 1984MNRAS.210..663M |
| [PS72]97 | 345.8481 | -26.2500 | sdA | 2023ApJ...950..141K |
| GD1532 | 347.5466 | -32.7012 | WD | TW |
| V*CGTuc | 352.2545 | -63.1107 | ApSi | 1975mcts.book.....H |

- Candidates that have been removed from the sample as they either possess poor astrometry or are too bright to be genuine hot subdwarfs (see Sec. 2.2.)
(TW) (This Work): Spectroscopically identified or reclassified in this work.